\documentclass{jcgt}

\usepackage{tikz}
\usepackage{pgfplots}
\usepackage{pgfplotstable}
\usepackage{subfigure}
\usetikzlibrary{spy}
\usetikzlibrary{shapes.geometric}
\usetikzlibrary{arrows, decorations.markings, backgrounds, fit, shadows, positioning}
\tikzstyle{box} = [draw, rectangle, rounded corners, ultra thick, node distance=7em, text width=10em, text centered, minimum height=7em, minimum width = 8em]
\tikzstyle{container} = [draw, rectangle, dashed, ultra thick, inner sep=0.5em]

\pgfdeclareimage{Teaser-SM}{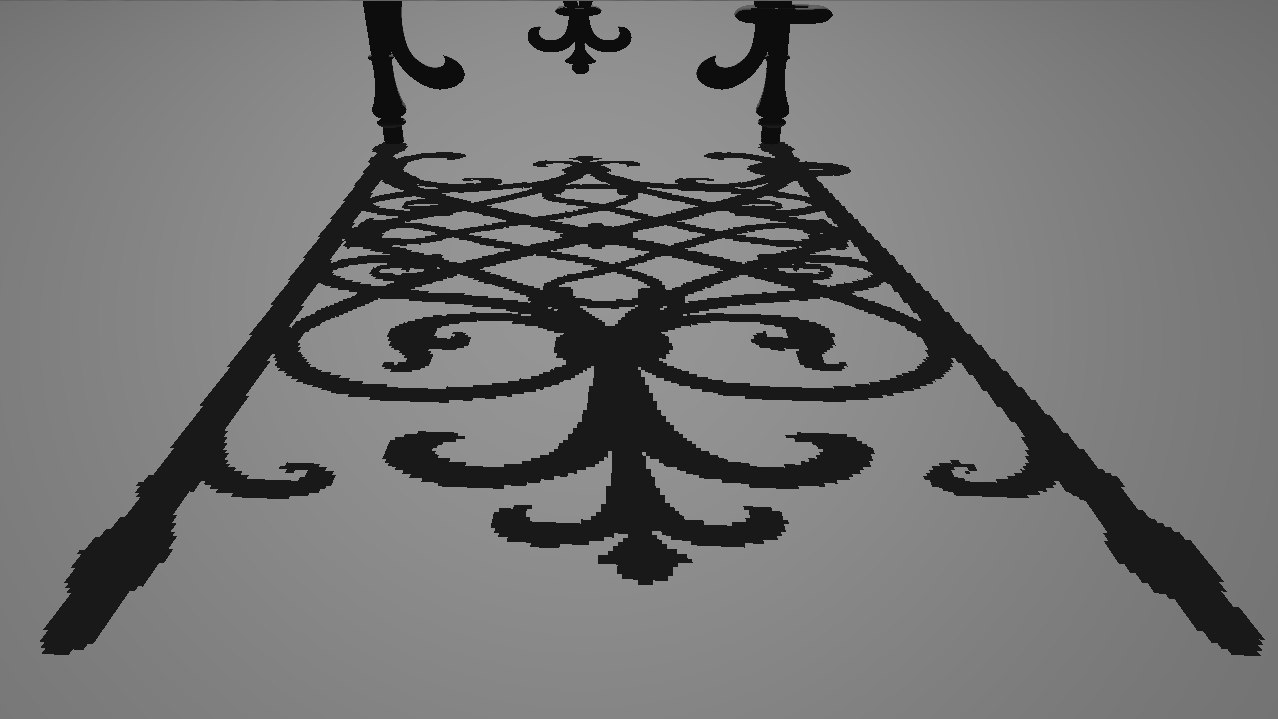}
\pgfdeclareimage{Teaser-SMSR}{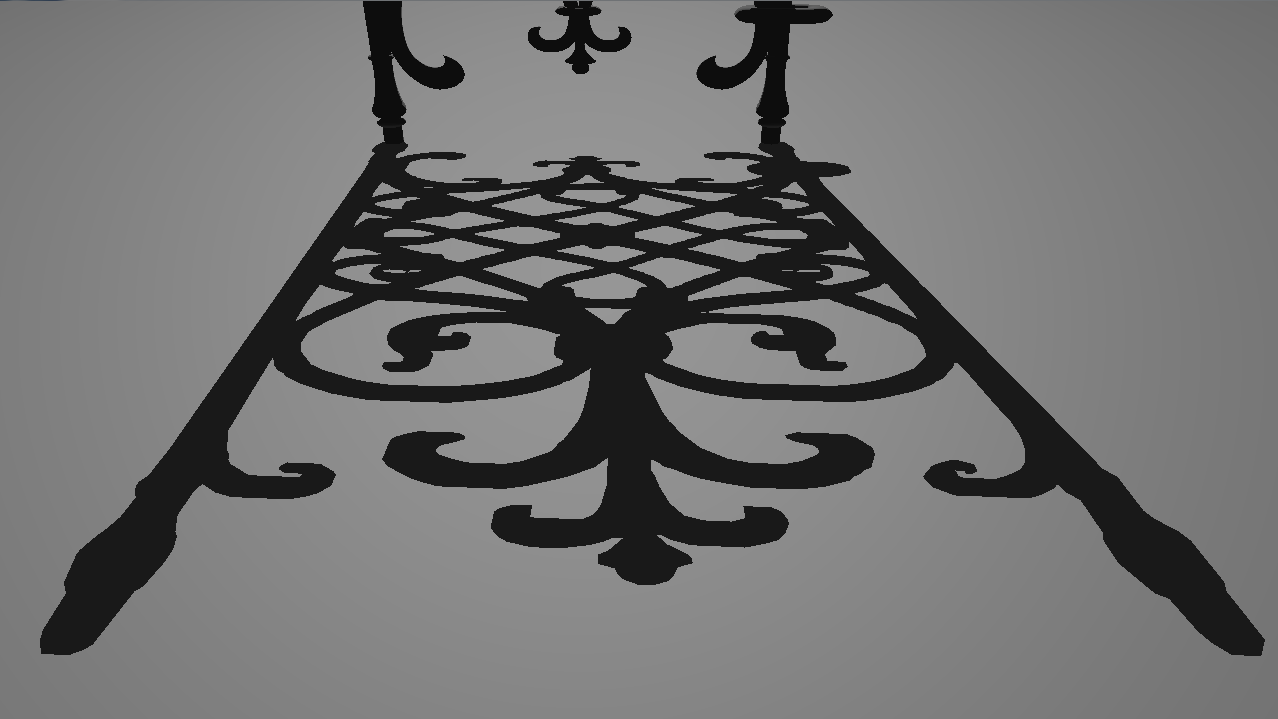}
\pgfdeclareimage{Teaser-RSMSS}{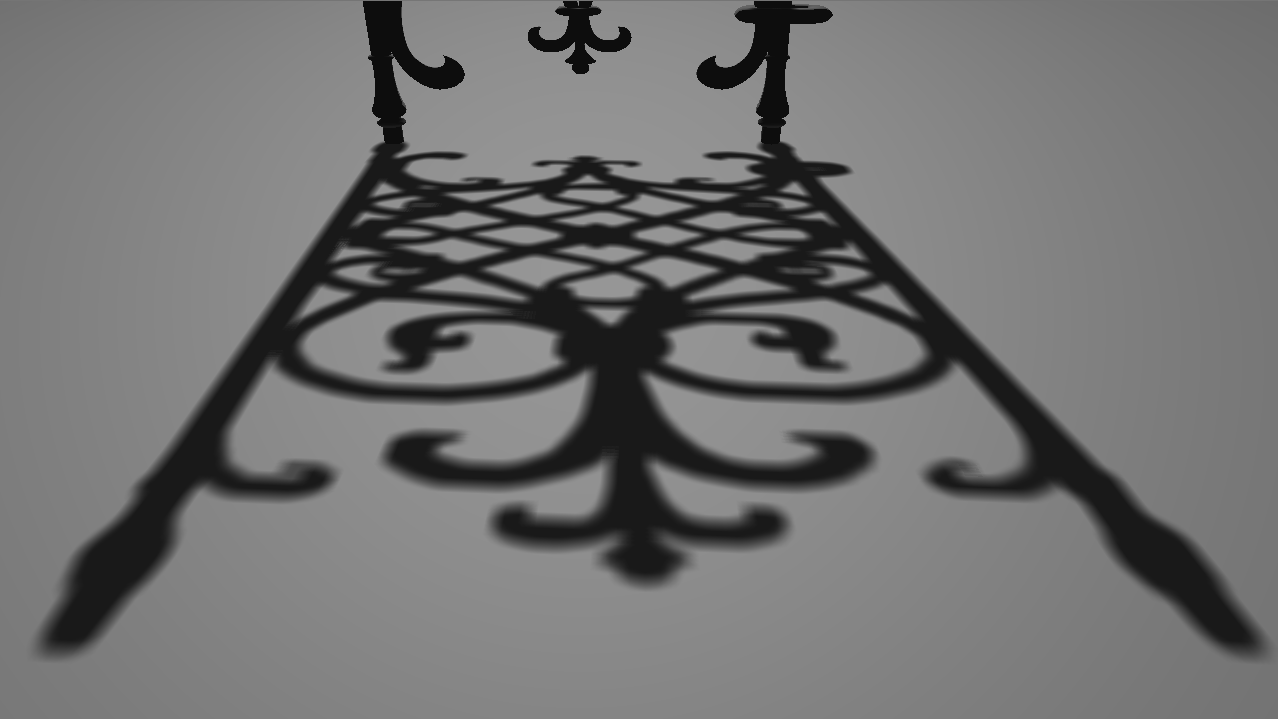}

\pgfdeclareimage{RBSM-SM}{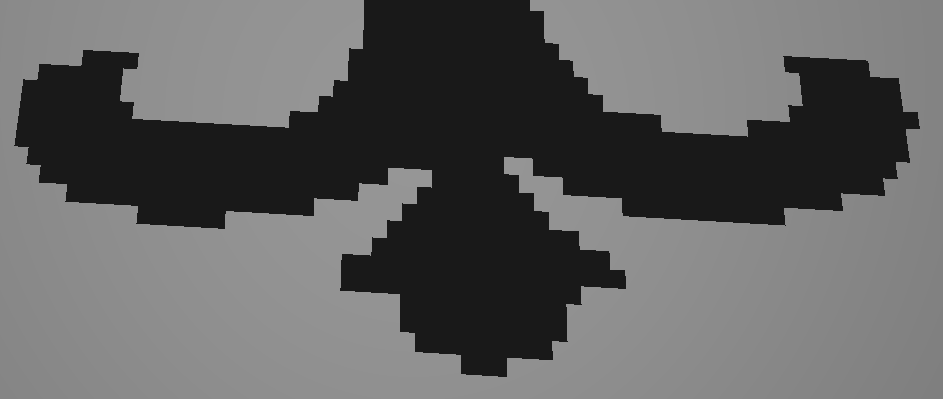}
\pgfdeclareimage{RBSM-SMSR}{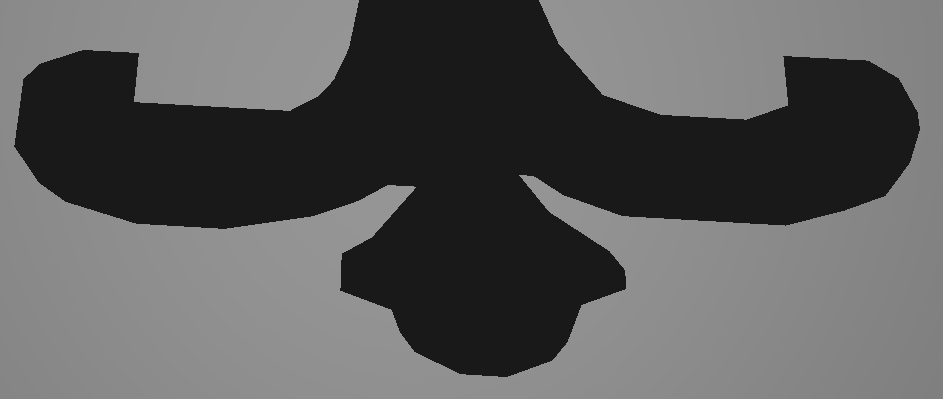}
\pgfdeclareimage{RBSM-SMSRDiscontinuity}{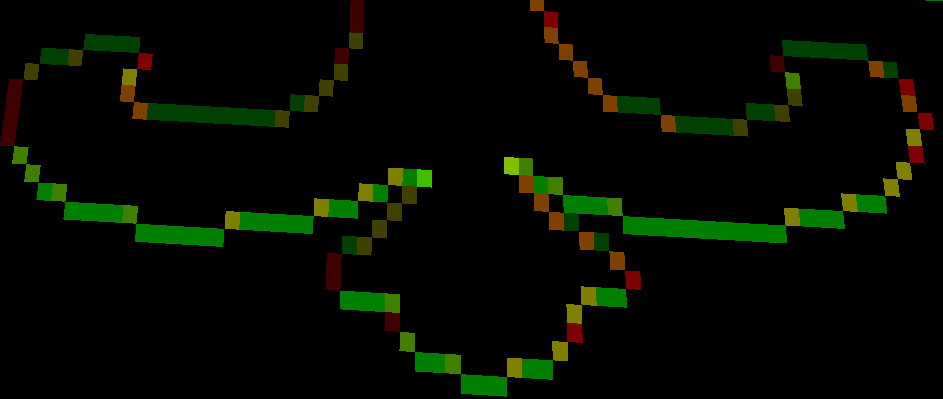}
\pgfdeclareimage{RBSM-SMSRONDS}{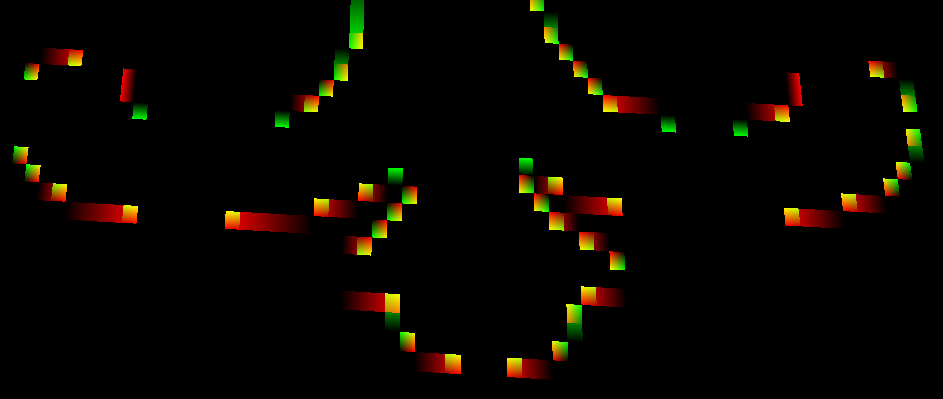}
\pgfdeclareimage{RBSM-SMSRClipping}{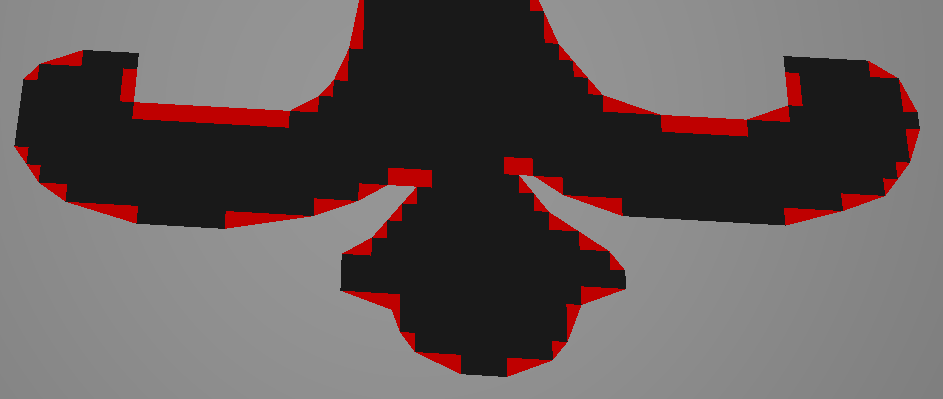}
\pgfdeclareimage{RBSM-RSMSS}{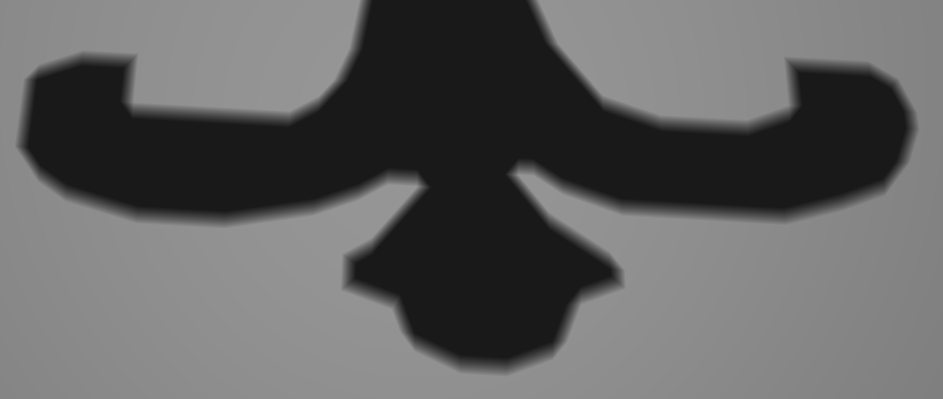}
\pgfdeclareimage{RBSM-RSMSSDiscontinuity}{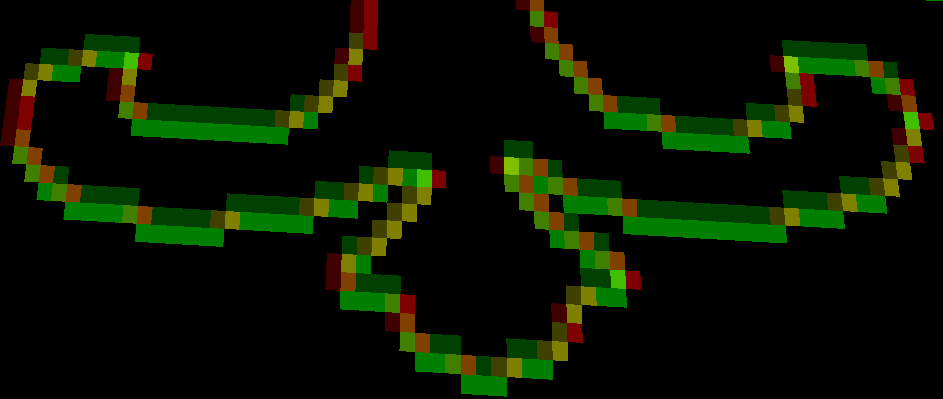}
\pgfdeclareimage{RBSM-RSMSSONDS}{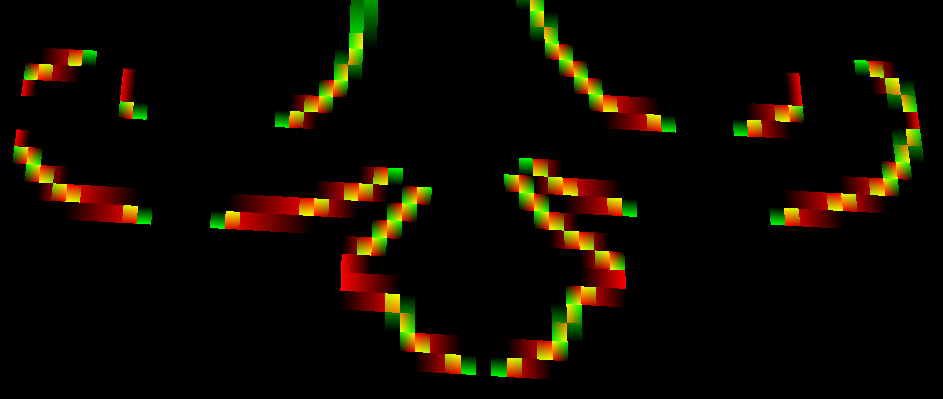}
\pgfdeclareimage{RBSM-RSMSSClipping}{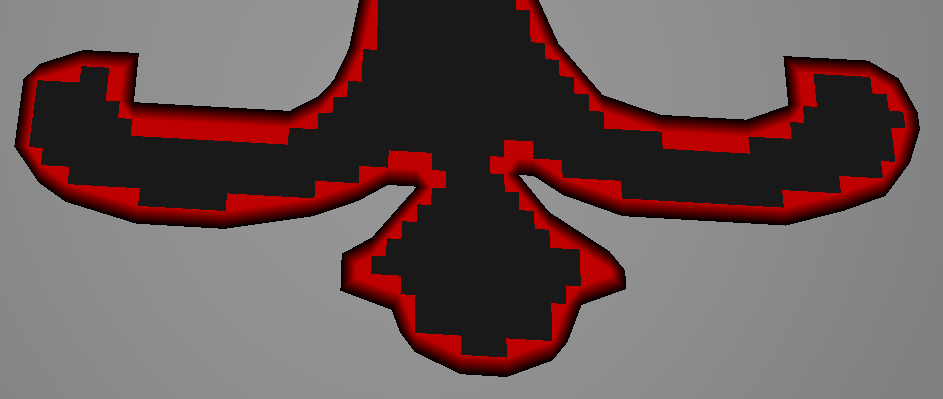}
\pgfdeclareimage{RBSM-RPCF}{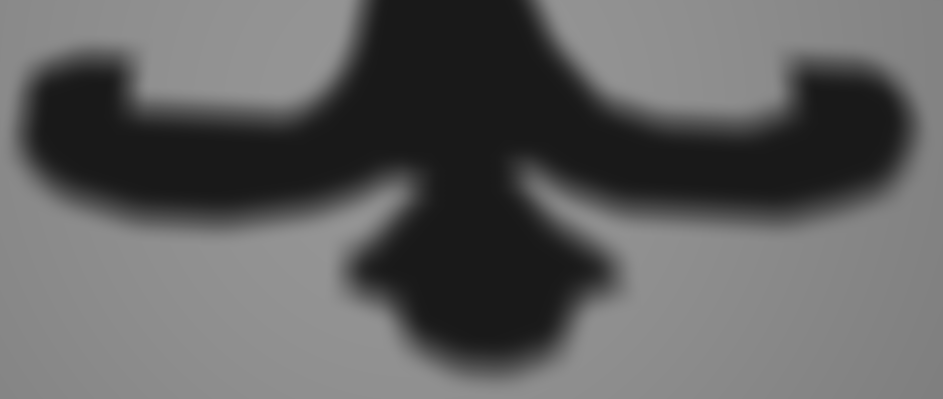}

\pgfdeclareimage{Limitations-SMSR-SM-1024}{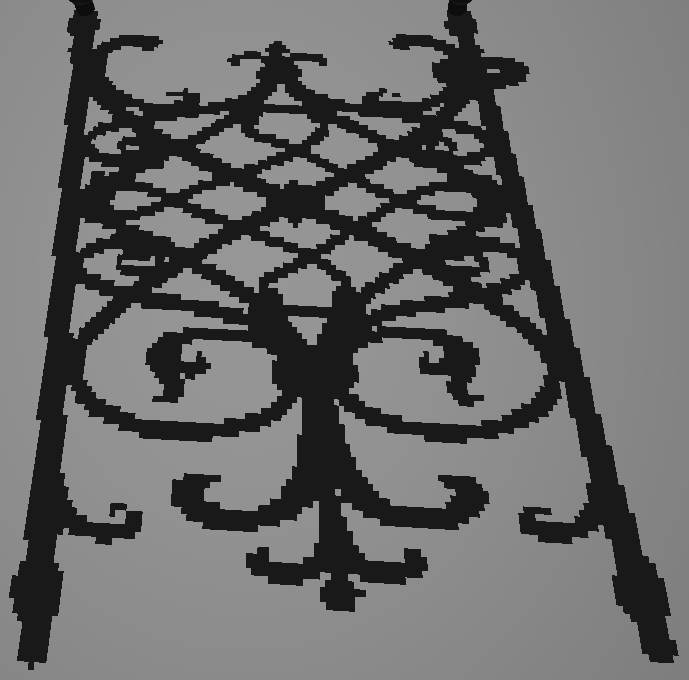}
\pgfdeclareimage{Limitations-SMSR-SM-2048}{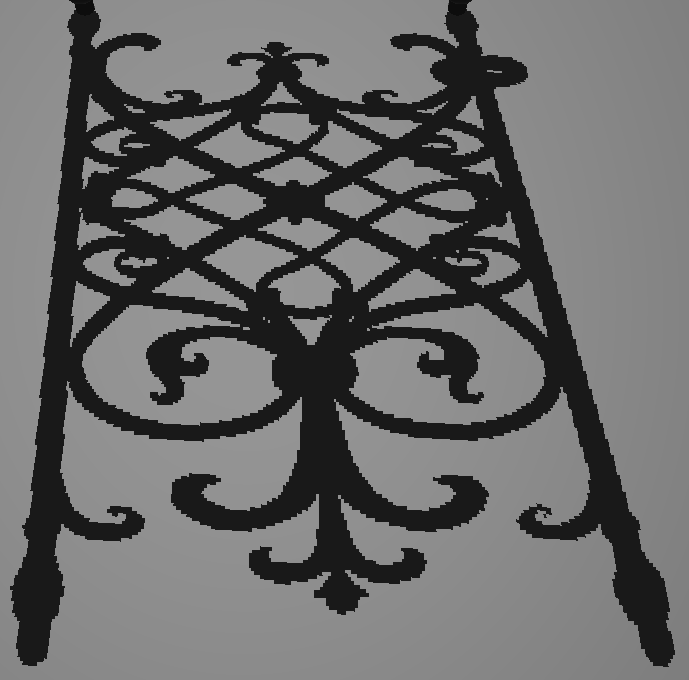}
\pgfdeclareimage{Limitations-SMSR-SM-4096}{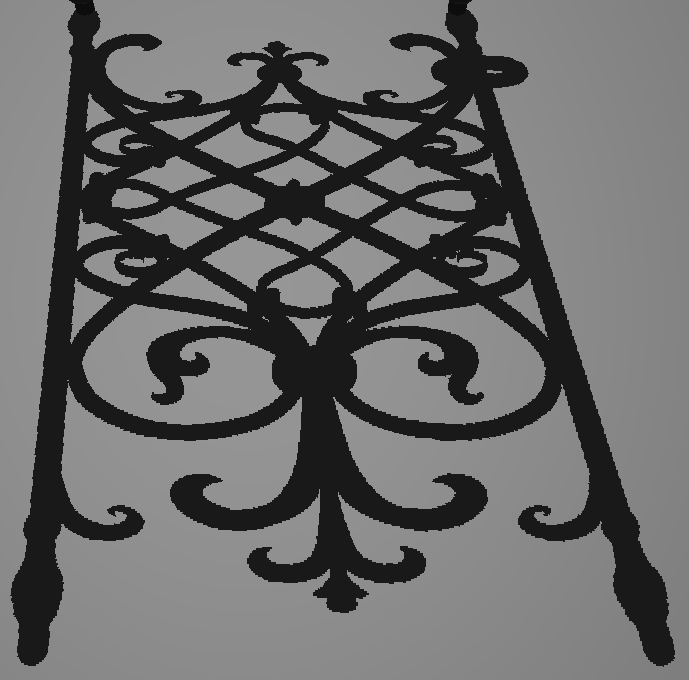}
\pgfdeclareimage{Limitations-SMSR-SMSR-1024}{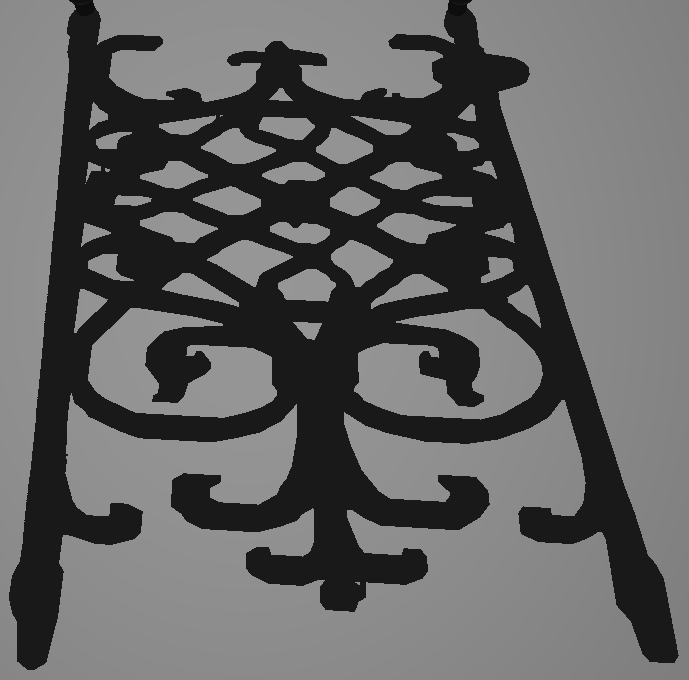}
\pgfdeclareimage{Limitations-SMSR-SMSR-2048}{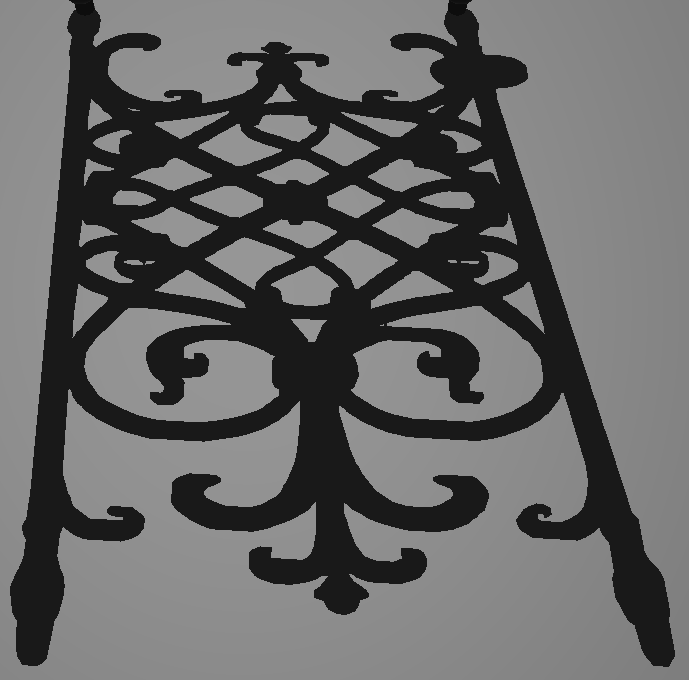}
\pgfdeclareimage{Limitations-SMSR-SMSR-4096}{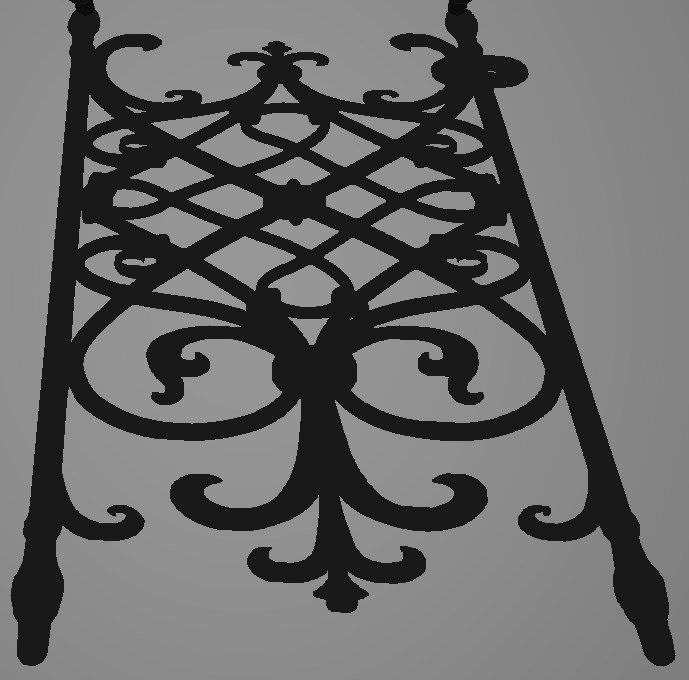}
\pgfdeclareimage{Limitations-RSMSS-SM}{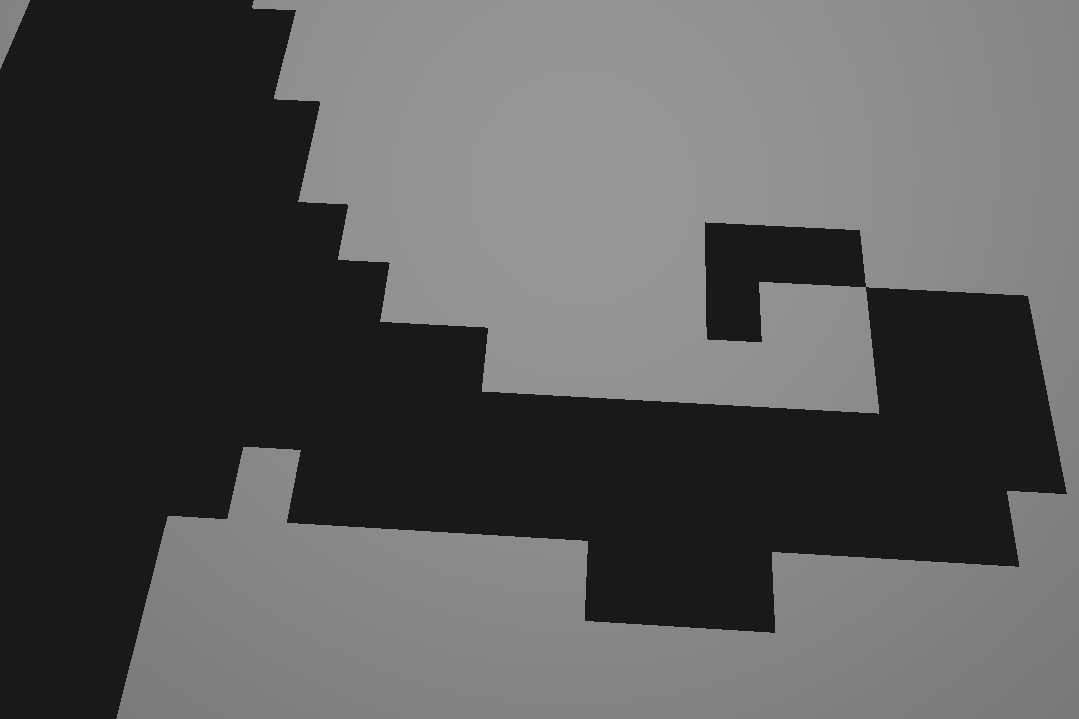}
\pgfdeclareimage{Limitations-RSMSS-RSMSS}{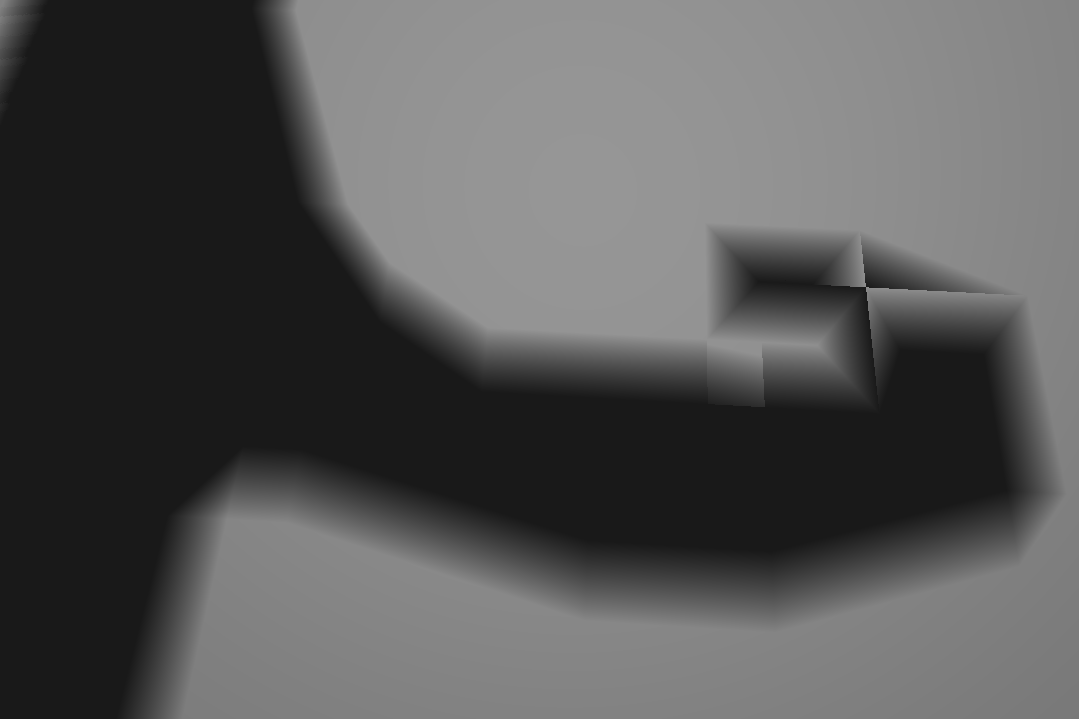}

\pgfdeclareimage{Fence-SM}{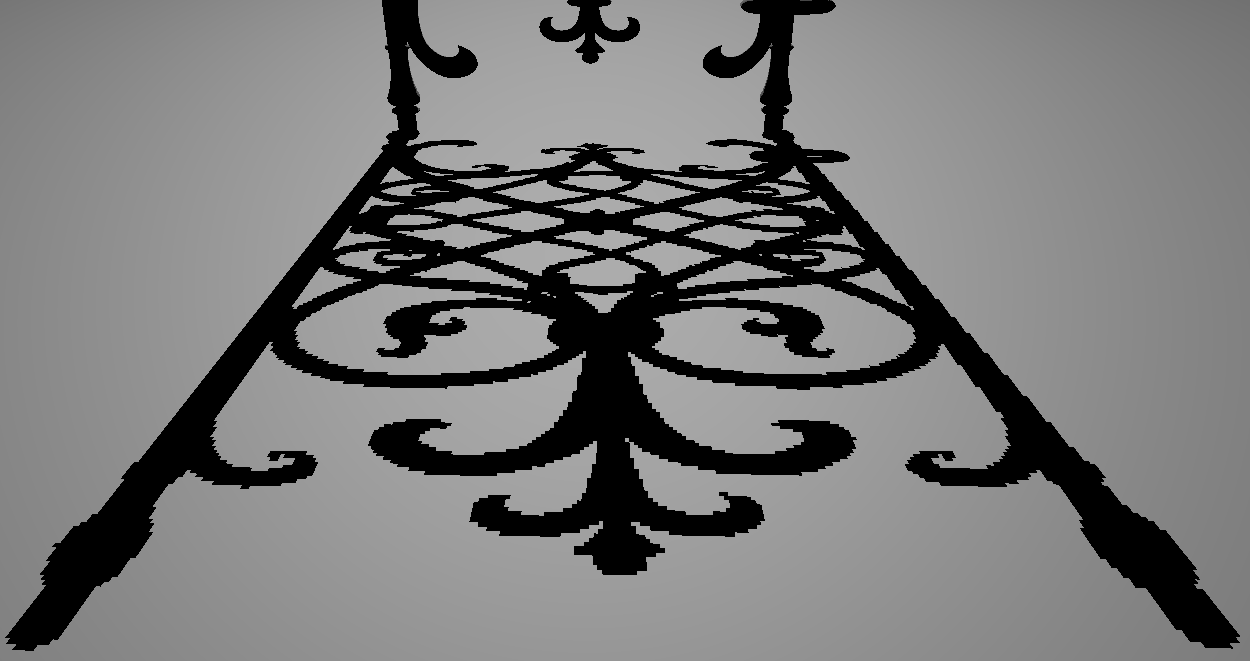}
\pgfdeclareimage{Fence-RBSM}{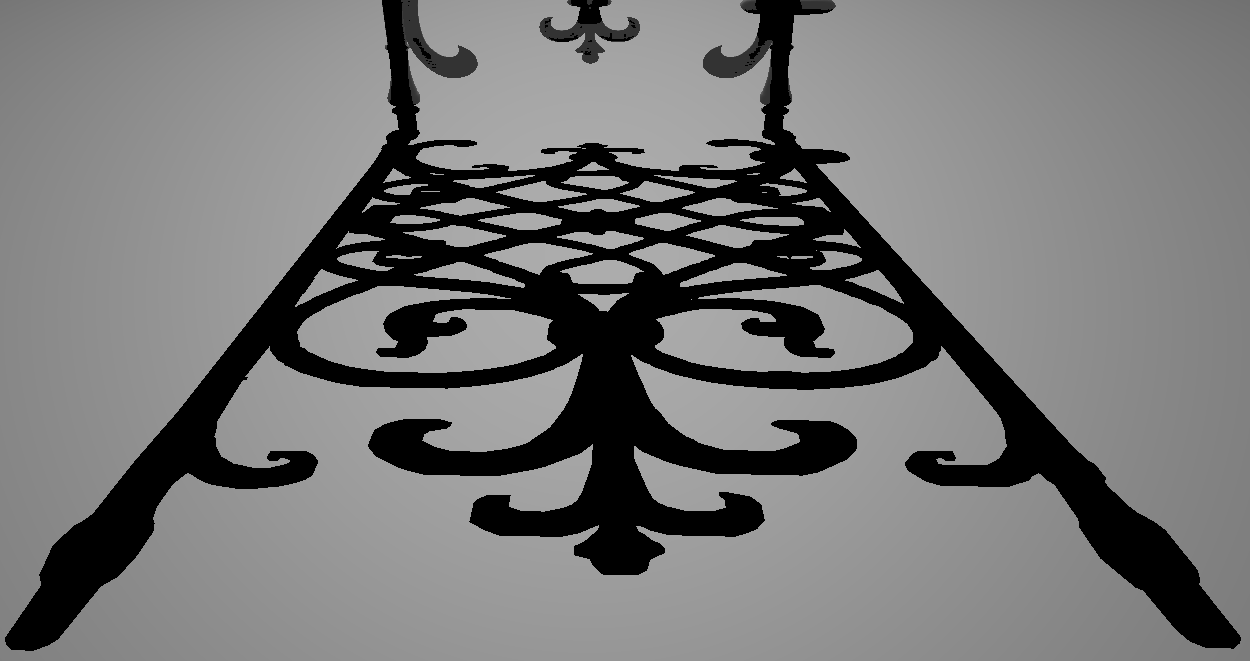}
\pgfdeclareimage{Fence-SV}{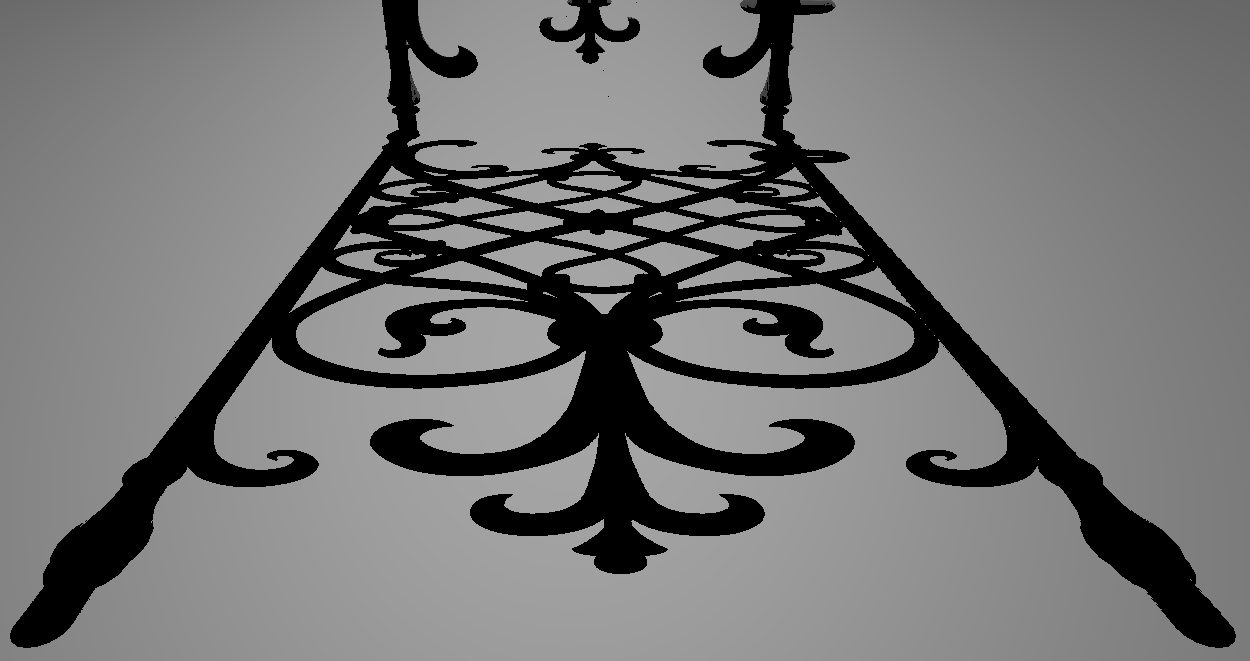}

\pgfdeclareimage{QuadBot-PCF}{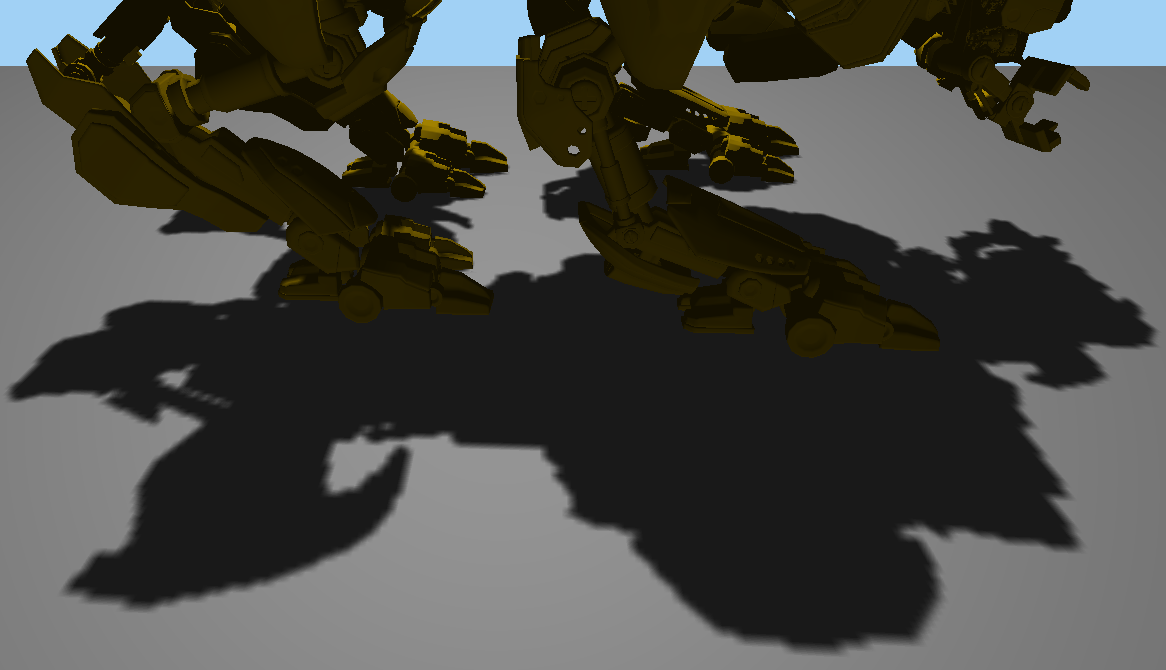}
\pgfdeclareimage{QuadBot-RBSM}{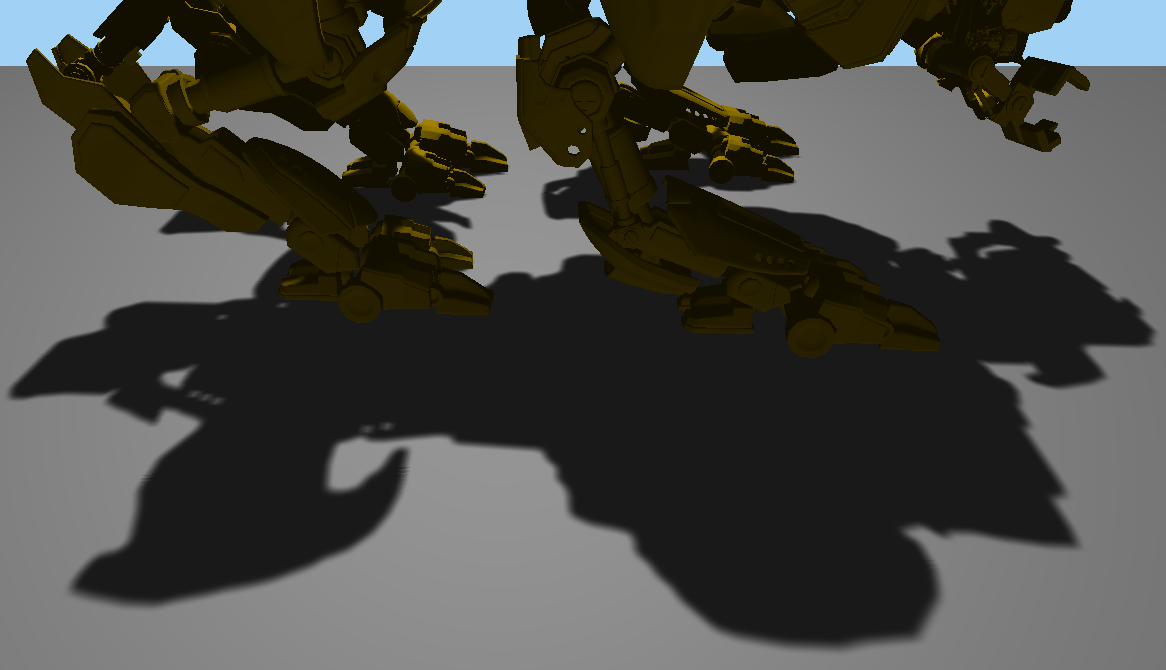}
\pgfdeclareimage{QuadBot-VSM}{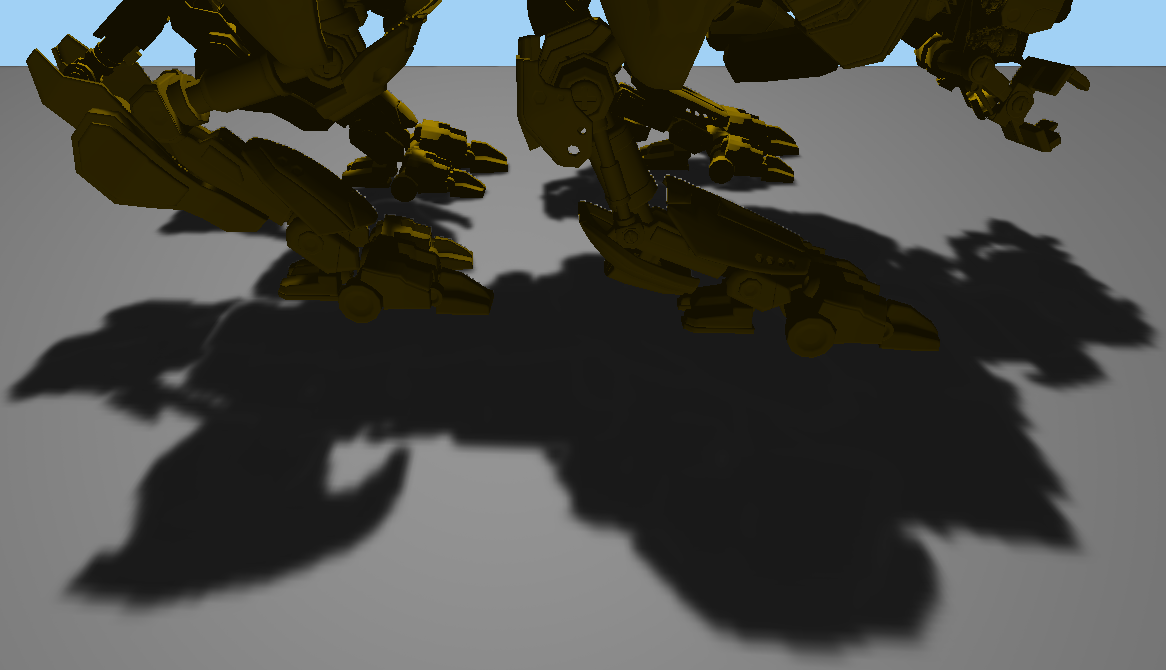}
\pgfdeclareimage{QuadBot-MSM}{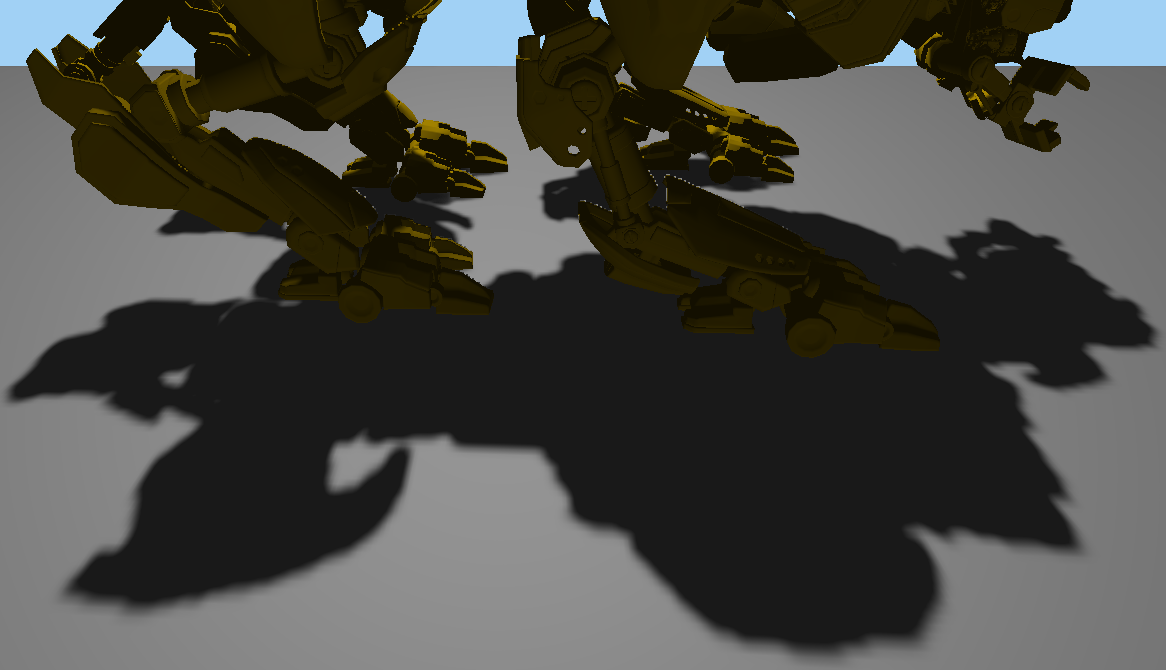}

\begin{document}

\title{Optimized Visibility Functions for Revectorization-Based Shadow Mapping}

\author
       {M\'arcio C. F. Macedo\\Federal University of Bahia
        \and Ant\^onio L. Apolin\'ario Jr.\\Federal University of Bahia
        \and Karl A. Ag\"uero \\Federal University of Bahia\thanks{TR-PGCOMP-007/2017. Technical Report. Computer Science Graduate Program. Federal University of Bahia.}
       }

\teaser{
  \centering
	\begin{tikzpicture}[>=stealth,line width=3, every node/.style={transform shape}, spy using outlines={rectangle,red,magnification=4.0, width=4.2cm, height=1.25cm}]
	
	\node [scale=0.155] at (0, 0) {\pgfuseimage{Teaser-SM}};
	\node [scale=0.155] at (4.4, 0) {\pgfuseimage{Teaser-SMSR}};
	\node [scale=0.155] at (8.8, 0) {\pgfuseimage{Teaser-RSMSS}};
			
	\spy [opacity=1.0] on (0.0, -0.6) in node [left, opacity=0.0] at (2.1, -2.0);
	\spy [opacity=1.0] on (4.4, -0.6) in node [left, opacity=0.0] at (6.5, -2.0);
	\spy [opacity=1.0] on (8.8, -0.6) in node [left, opacity=0.0] at (10.9, -2.0);
	
	\end{tikzpicture}
	\caption{Given the jagged shadow edges generated with shadow mapping (left), revectorization-based shadow mapping is able to reduce aliasing (center) and simulate penumbra (right). In this paper, we propose a new, optimized set of visibility functions to ease the implementation of the shadow revectorization technique, while improving its performance.} 
  \label{fig:teaser}
}

\maketitle

\begin{abstract}
\small
High-quality shadow anti-aliasing is a challenging problem in shadow mapping. Revectorizati-on-based shadow mapping (RBSM) minimizes shadow aliasing by revectorizing the jagged shadow edges generated with shadow mapping, keeping low memory footprint and real-time performance for the shadow computation. However, the current implementation of RBSM is not so well optimized because its visibility functions are composed of a set of 43 cases, each one of them handling a specific revectorization scenario and being implemented as a specific branch in the shader. Here, we take advantage of the shadow shape patterns to reformulate the RBSM visibility functions, simplifying the implementation of the technique and further providing an optimized version of the RBSM. Our results indicate that our implementation runs faster than the original implementation of RBSM, while keeping its same visual quality and memory consumption. Furthermore, we show GLSL source codes to ease the implementation of our technique, provide a comparison between the optimized RBSM and related work, and discuss the limitations of the shadow revectorization.
\end{abstract}

\section{Introduction}
\label{sec:introduction}

Shadows are important because they provide photorealism, enhancing our understanding of a scene and improving our comprehension about the spatial relationships between light blocker and shadow receiver objects. Games and augmented reality applications make use of shadows to improve the visual quality of their computer-generated scenes. In these applications, shadows must be computed in real-time (to keep user interactivity) and with high visual quality (to improve the user's perception of the virtual scene).

Shadow mapping \cite{Williams1978} is one of the fastest methods used to compute shadows in real-time. However, the image-based representation of the technique generates shadows prone to aliasing artifacts along the shadow edge (Figure \ref{fig:teaser}-left) and temporal incoherency as well.

To generate anti-aliased shadows in real-time, revectorization-based shadow mapping (RBSM) \cite{Macedo2016} proposes the revectorization (Figure \ref{fig:teaser}-center) and filtering (Figure \ref{fig:teaser}-right) of the jagged shadow edges generated with shadow mapping. Indeed, RBSM is able to improve the accuracy of the shadow mapping at little additional cost. However, the definition of the RBSM visibility functions is composed of 43 cases, which must be explicitly detected by branches in the shader code. In this paper, we want to exploit the jagged shadow edge shape pattern as well as the awareness of symmetric cases to simplify the RBSM visibility functions, keeping high visual quality, improving performance, and easing the implementation of the technique. 

Our main contributions are threefold:

\begin{enumerate}

\item A new, compact visibility function which takes advantage of the shadow shape pattern and symmetric cases to speed up the shadow revectorization;

\item An optimized visibility function for revectorization-based shadow filtering;

\item An in-depth discussion of the shadow revectorization technique, including implementation details;

\end{enumerate}

The remainder of this work is organized as follows: Section \ref{sec:relatedwork} covers the relevant work in the field of real-time shadows. Section \ref{sec:RBSM} reviews the RBSM technique. In Section \ref{sec:VisibilityFunctions}, we present our proposal of optimized visibility functions for RBSM. Section \ref{sec:Results} shows comparative results between the optimized RBSM and related work in terms of visual quality and rendering time. A summary of the paper, as well as plans for future work can be seen in Section \ref{sec:Conclusion}.

\section{Related Work}
\label{sec:relatedwork}

In the shadow mapping technique, the depth buffer of the scene seen from the light source viewpoint is stored in the shadow map, whose values are compared to the depth values of the scene seen from the camera viewpoint to determine the visibility condition of each fragment present in the scene. As discussed in Section \ref{sec:introduction}, the finite resolution of the shadow map generates aliasing artifacts along the shadow edge, which lower the shadow visual quality. 

Several strategies have been used to extend the shadow mapping technique to generate anti-aliased shadows in real-time. This section covers the most relevant approaches shown in the literature. A more complete review of related work can be found in \cite{Eisemann2011,Woo2012}.

\paragraph{Warping: } Aliasing artifacts can be reduced by the reparametrization of the shadow map generation. In other words, by changing the mapping function that transforms the world-space coordinates to the shadow map texture coordinates, one can improve the shadow map resolution in the region near the viewpoint, while lowering the sampling density in the regions far from the viewpoint. Logarithmic parametrization \cite{Lloyd2008} is the technique which allows the higher accurate warping, at the cost of lower frame rate.

\paragraph{Partitioning: } An alternative to reduce the aliasing artifacts caused by the use of an insufficient shadow map resolution is to make use of several shadow maps generated from different locations to better sample the depth of the objects located near and far the camera viewpoint \cite{Lefohn2007,Lauritzen2011}.

Partitioning techniques are useful to reduce aliasing artifacts caused mainly by the use of a single shadow map in a large-scale virtual environment. To further improve the accuracy of the solution, these techniques are commonly associated with warping techniques. Thus, the several shadow maps generated from the partitioning approaches are built using an improved reparametrization approach. While this combination may work in some situations, for others, the use of partitioned shadow maps only increases memory usage and processing time, and does not guarantee a really improved visual quality.

\begin{figure}[ht]
\centering
  \includegraphics[width=0.975\linewidth]{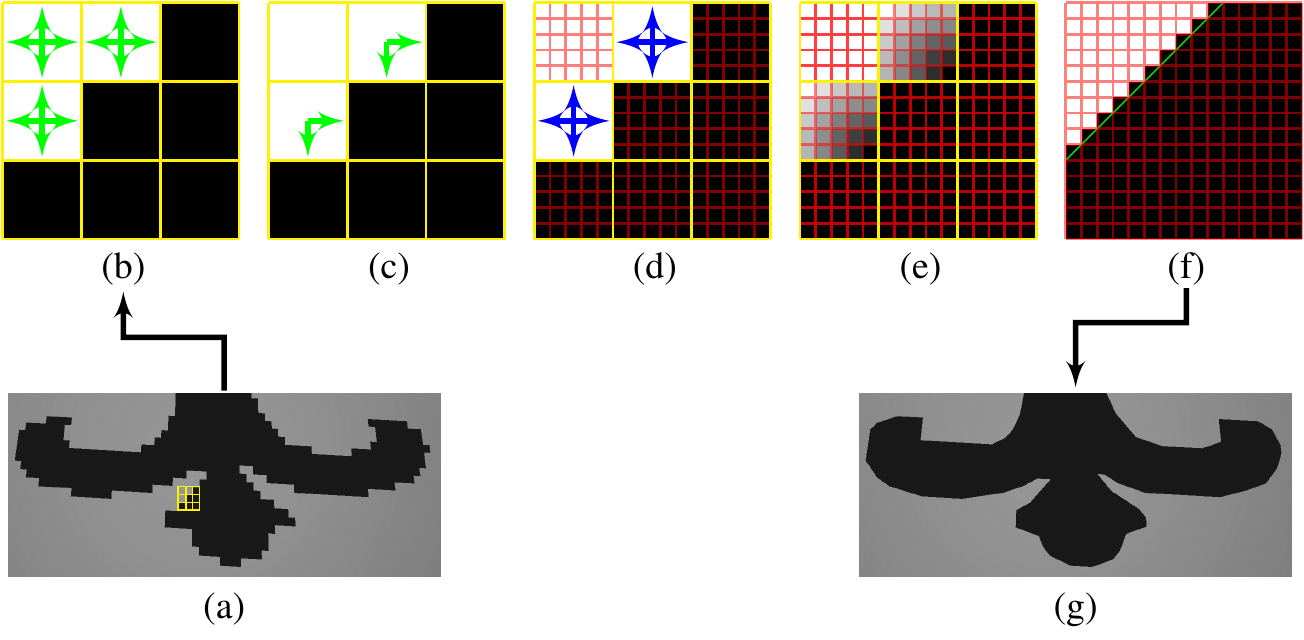}
  \caption{An overview of the RBSM pipeline for shadow silhouette recovery. Given an aliased shadow edge (a), RBSM operates over the lit fragments to evaluate the spatial coherency between shadow tests (b). Then, RBSM detects the directions (green arrows) of where the shadow silhouette is located (c), estimates the size of the aliased edge (d) and computes the normalized distance of each camera-view fragment (red grid) to the origin of the aliased edge (e). Finally, the algorithm traces a revectorization line (green line in (f)) to determine the final shadow intensity of each fragment (f, g) in the camera space.} 
  \label{fig:SMSR}
\end{figure}

\paragraph{Filtering: } To simulate the penumbra effect and further suppress aliasing artifacts, filtering techniques smooth the jagged shadow edges by the use of low- or high-order kernels. Traditional techniques \cite{Reeves1987} provide good anti-aliasing, but suffer from scalability issues, while the state-of-the-art ones \cite{Annen2008,Peters2015} are based on shadow map pre-filtering, providing scalable and real-time performance, but generating light leaking artifacts (\textit{i.e.}, where a shadowed fragment is incorrectly rendered as a lit one).

Filtering techniques solve the problems of anti-aliasing and penumbra simulation in real-time. However, for low-resolution shadow maps, low-order filter sizes produce blurred jagged shadow edges. High-order filter sizes remove the jagged appearance of the shadow edges, but may smooth out fine details along the shadow edge.

\paragraph{Silhouette Recovery: } Some techniques aim to compute accurate shadow silhouettes either by proposing a hybrid approach \cite{Hertel2009} or embedding additional geometric data into the shadow map representation \cite{Lecocq2014}. In both cases, accurate shadows are generated at the cost of increased processing time and high memory consumption. 

Inspired by the shadow map silhouette revectorization \cite{Bondarev2014}, RBSM is a technique which is able to generate accurate shadows by using either filtering or silhouette recovery strategies. The technique keeps low memory consumption and processing time of the shadow mapping because it works directly on the shadow edges generated by shadow mapping. Even in this case, we show in the remaining of this paper that RBSM can be more efficient, mainly by the reformulation of its visibility functions.

\section{Revectorization-Based Shadow Mapping}
\label{sec:RBSM}

In this section, we present both theoretical and implementation details of RBSM. To do so, we first present the RBSM, which provides shadow silhouette recovery, and whose overview is shown in Figure \ref{fig:SMSR}. Then, we present the variant of RBSM which provides anti-aliased shadow filtering. Both variants are implemented in a single-pass in the shader, as shown in Listing \ref{lst:RBSM}, without the use of any additional textures, besides the shadow map. A discussion about the optimized RBSM visibility functions is shown in the next section, where we present our proposal. 

\begin{lstlisting}[caption={GLSL code for RBSM implementation.}, label={lst:RBSM}]
uniform sampler2D SM; //Shadow map texture
uniform int SMWIDTH; //Shadow map width
uniform int SMHEIGHT; //Shadow map height
uniform bool silhouetteRecovery = true;
varying vec3 p; //Fragment transformed into light space coordinates

float RBSM(vec3 p) {

	//Retrieve the depth of the blocker of p
	float z = texture2D(SM, p.xy).z;
	//Compute the shadow test
	float shadowTestResult = shadowTest(p.z, z);
	//For silhouette recovery only
	if(silhouetteRecovery) {
		//Discard shadowed fragments from further RBSM computation
		if(shadowTestResult == 0) return shadowTestResult;
	}
	//Compute the shadow map offset
	vec2 o = vec2(1.0/SMWIDTH, 1.0/SMHEIGHT);
	//Compute the discontinuity
	vec3 d = computeDiscontinuity(p, z, o);
	//If the fragment is located in the shadow edge
	if(d.r > 0.0 || d.g > 0.0) {
		//Compute the relative distance of p (rd) to the shadow edge
		vec4 rd = computeFragmentDistanceToSE(p, d, o);
		//Normalize the relative distance to the unit interval
		vec4 nrd = normalizeFragmentDistanceToSE(rd);
		//Revectorize the shadow edge
		return computeRBSMVisibility(d, nrd);
	//Else, return the shadow test result
	} else return shadowTestResult;
	
}

\end{lstlisting}

\subsection{Revectorization-Based Shadow Silhouette Recovery}

\subsubsection{Overview}

The RBSM algorithm aims to locate shadow edge patterns in the scene (Figure \ref{fig:SMSR}-(a)) and to use the available screen-space resolution provided by the camera view to perform shadow anti-aliasing through the revectorization of the shadow (Figure \ref{fig:SMSR}-(f)).

As shown in Figure \ref{fig:SMSR}-(b), the first step of RBSM consists of an evaluation of the difference of shadow test results between neighbour shadow map texels. The goal of this step is to detect where the shadow aliasing is located. This step is performed only for lit fragments (Lines 14-17 of Listing \ref{lst:RBSM}) because the revectorization-based shadow silhouette recovery aims to minimize shadow aliasing by working over the lit-side of the shadow edge. Hence, since shadowed fragments will remain in shadow after the silhouette recovery, they are discarded from the additional computation required by RBSM. After the evaluation of the spatial coherency between neighbour shadow test results, the algorithm is able to detect the discontinuity directions (green arrows in Figure \ref{fig:SMSR}-(c)) where the jagged shadow edges, or shadow discontinuities, are located
(Figure \ref{fig:SMSR}-(c) and Lines 18-21 of Listing \ref{lst:RBSM}).

For each fragment inside a shadow edge (Lines 22-23 of Listing \ref{lst:RBSM}), the algorithm performs a traversal over the shadow edge in order to compute the size of the shadow edge, as well as the relative distance and position of each fragment with respect to the end of the shadow edge (Figure \ref{fig:SMSR}-(d) and Lines 24-25 of Listing \ref{lst:RBSM}). Next, the algorithm normalizes the distance and position of each fragment to the shadow edge (Figure \ref{fig:SMSR}-(e) and Lines 26-27 of Listing \ref{lst:RBSM}), such that a set of linear comparisons can be used to define a revectorization line (green line in Figure \ref{fig:SMSR}-(f)), which determines whether a fragment must be shadowed by RBSM (Figures \ref{fig:SMSR}-(f, g) and Lines 26-29 of Listing \ref{lst:RBSM}). Meanwhile, the shadow intensity of fragments outside the shadow edge is determined by the traditional shadow test (Lines 30-31 of Listing \ref{lst:RBSM}).

In this subsection, we have presented an overview of the revectorization-based
shadow silhouette recovery. A detailed description of the step by step of this algorithm can be seen in the following subsections.

\begin{lstlisting}[caption={GLSL code for discontinuity computation.}, label={lst:DiscontinuityComputation}, float]
vec3 computeDiscontinuity(vec3 p, float z, vec2 o) {

	//Perform the shadow test for the current fragment
	float s = shadowTest(p.z, z);
	//Perform the shadow test for the 4-connected neighbourhood
	vec4 N = computeN(p, o);
	//Compute the discontinuity directions
	vec4 dU = abs(N - s);
	//Store discontinuity directions (dC.xy) + shadow test (dC.z)
	vec2 dC = (2.0 * dU.xz + dU.yw)/4.0;
	return vec3(dC, 1.0 - s);
	
}

vec4 computeN(vec3 p, vec2 o) {

	vec4 N;
	N.x = shadowTest(p.z, texture2D(SM, vec2(p.x - o.x, p.y)).z));
	N.y = shadowTest(p.z, texture2D(SM, vec2(p.x + o.x, p.y)).z));
	N.z = shadowTest(p.z, texture2D(SM, vec2(p.x, p.y + o.y)).z));
	N.w = shadowTest(p.z, texture2D(SM, vec2(p.x, p.y - o.y)).z));
	return N;
	
}
\end{lstlisting}

\subsubsection{Shadow Edge Localization}

Shadow edges are detected according to the difference between the illumination condition of the neighbour shadow map texels (Listing \ref{lst:DiscontinuityComputation}).

Let us assume a point \textbf{p} whose distance to the light source is defined by \textbf{p}$_{z}$. If we denote the shadow map texel \textbf{t}$_{x,y}$ located at the 2D position $x, y$ and the function $z(\textbf{t}_{x,y})$ as a function that computes the depth of the blocker of \textbf{p} stored in the corresponding shadow map texel \textbf{t}$_{x,y}$, we can define the binary shadow test visibility function $s(\textbf{p}_{z}, z(\textbf{t}_{x,y}))$ (function \lstinline{shadowTest} in Line 12 of Listing \ref{lst:RBSM} and Lines 4, 18-21 of Listing \ref{lst:DiscontinuityComputation}) as \cite{Williams1978}

\begin{equation}
\label{eq:ShadowTest}
\textit{s}(\textbf{p}_{z}, \textit{z}(\textbf{t}_{x, y})) = 
\begin{cases}
	0 & \text{if } \textbf{p}_{z} >  \textit{z}(\textbf{t}_{x, y}),
	\\
	1 & \text{otherwise}.
\end{cases}
\end{equation}

The shadow test (\ref{eq:ShadowTest}) indicates that the point \textbf{p} is in shadow $(s(\textbf{p}_{z}, z(\textbf{t}_{x,y})) = 0)$ if the blocker of \textbf{p} (with depth $z(\textbf{t}_{x,y})$) is closer to the light source than \textbf{p} (with depth \textbf{p}$_{z}$).

Given the shadow test defined in (\ref{eq:ShadowTest}), the first step to detect shadow edges consists on the computation of (\ref{eq:ShadowTest}) to determine the illumination condition of each fragment visible in the scene (Figure \ref{fig:SMSR}-(a) and Lines 3-4 of Listing \ref{lst:DiscontinuityComputation}). Then, the difference between shadow tests of the current fragment and its 4-connected neighbours in the shadow map (Figure \ref{fig:SMSR}-(b)) is estimated as N (function \lstinline{computeN} in Lines 15-24 of Listing \ref{lst:DiscontinuityComputation}).

\begin{equation}
\label{eq:NeighbourhoodEvaluation}
\textbf{N} = \begin{bmatrix}
\textit{s}(\textit{z}(\textbf{t}_{x - o_{x}, y})), \textit{s}(\textit{z}(\textbf{t}_{x + o_{x}, y})), \textit{s}(\textit{z}(\textbf{t}_{x, y + o_{y}})), \textit{s}(\textit{z}(\textbf{t}_{x, y - o_{y}}))
\end{bmatrix},
\end{equation}
where $o_{x}$ and $o_{y}$ are the shadow map offset values (Lines 18-19 of Listing \ref{lst:RBSM}) and \textbf{p}$_{z}$ was omitted from (\ref{eq:NeighbourhoodEvaluation}) for clarity, because \textbf{p}$_{z}$ has the same value for the four shadow test evaluations.

Once the neighbourhood evaluation is computed, we can detect the directions
where the shadow edges are located. The RBSM uses the concept of discontinuity
(green arrows in Figure \ref{fig:SMSR}-(c), Lines 7-8 in Listing \ref{lst:DiscontinuityComputation}) to store whether a fragment is located in the inner- or the outer-side of the shadow edge, and where the shadow edge is located with respect to its 4-connected neighbourhood in the shadow map. Discontinuity can be simply computed as the absolute difference in the shadow test results between a fragment and its 4-connected neighbors in the shadow map

\begin{equation}
\label{eq:Discontinuity}
\textbf{d} = \begin{bmatrix}
|\textbf{N}_{1} - s|, |\textbf{N}_{2} - s|, |\textbf{N}_{3} - s|, |\textbf{N}_{4} - s|\end{bmatrix}.
\end{equation}

\begin{table}[b]
\centering
\begin{tabular}{|c|c|c|c|}
\hline
\multicolumn{1}{|l|}{} & \multicolumn{3}{c|}{\textbf{Discontinuity Components}} \\ \hline
\textbf{Value}         & d.x      & d.y      & d.z\\ \hline
0              & No edge   & No edge   & Lit fragment     \\ \hline
0.25           & Right-side edge              & Top-side edge               & --           \\ \hline
0.5            & Left-side edge               & Bottom-side edge            & --           \\ \hline
0.75           & Left-right-side edge     & Top-bottom-side edge    & --      		\\ \hline
1              & --     		    & --     			 & Shadowed fragment 		\\ \hline\end{tabular}
\caption{The meaning of the values stored for each discontinuity component.}
\label{tab:Discontinuity}
\end{table}

Discontinuity \lstinline{d}, as defined in (\ref{eq:Discontinuity}), represents a \lstinline{vec4} (\lstinline{d} $\in [0; 1]$) that stores whether a shadow edge exists for a particular direction. For instance, the first position of \lstinline{d}, indexed as \lstinline{d.x} in Listing \ref{lst:DiscontinuityComputation}, stores whether a shadow edge exists (\lstinline{d.x = 1}) or not (\lstinline{d.x = 0}) at the left side of the current fragment. The second position \lstinline{d.y}, stores whether a shadow edge exists on the right side of the current fragment. The
third and fourth positions of \lstinline{d}, \lstinline{d.z} and \lstinline{d.w}, store information related to the top and bottom sides, respectively.

To keep consistency with previous definitions of discontinuity \cite{Bondarev2014,Macedo2016}, we compute an alternative version of the discontinuity (Lines 9-11 in Listing \ref{lst:DiscontinuityComputation})

\newsavebox\boxX
\newsavebox\boxY
\newsavebox\boxZ
\newsavebox\boxW
\savebox\boxX{\lstinline{d.x}}
\savebox\boxY{\lstinline{d.y}}
\savebox\boxZ{\lstinline{d.z}}
\savebox\boxW{\lstinline{d.w}}

\begin{eqnarray}\nonumber
\label{eq:compressedDiscontinuity}
\usebox{\boxX} &=& \frac{2\usebox{\boxX} + \usebox{\boxY}}{4}\\
\usebox{\boxY} &=& \frac{2\usebox{\boxZ} + \usebox{\boxW}}{4}\\
\usebox{\boxZ} &=& 1 - s(\textbf{p}_{z}, z(\textbf{t}_{x,y})),\nonumber
\end{eqnarray}
which stores the directions where the shadow edge is located along the horizontal (\lstinline{d.x}) and vertical axes (\lstinline{d.y}) in the first two components of the discontinuity and whether the fragment is located in the inner or the outer side of the shadow edge in the third component of the discontinuity \lstinline{d.z}. Table \ref{tab:Discontinuity} shows the mapping between the values and shadow edge directions represented by the discontinuity.

With the computation of (\ref{eq:compressedDiscontinuity}), we are able to detect where the shadow edges are located and store their directions on the basis of the discontinuity representation. 

\begin{lstlisting}[caption={GLSL code for shadow edge traversal.}, label={lst:DiscontinuitySpaceTraversal}, float]
uniform int MAXDIST = 16; //Maximum shadow edge size

vec2 traverseShadowSilhouette(vec3 p, vec2 d, vec2 dir, vec2 o) {

	float shadowEdgeEnd = 0.0;
	float distance = 1.0;
	bool hasDiscontinuity = false;
	vec3 np = p;
	//Compute the step of traversal for current direction
	vec2 step = dir * o;
	
	//Iteratively traverse the shadow silhouette
	for(int i = 0; it < MAXDIST; it++) {
		//Access the neighbour of p
		np.xy += step;
		//Retrieve the depth of the blocker of np
		float z = texture2D(SM, np.xy).z;
		//Determine the visibility of np
		float s = shadowTest(np.z, z);
		//When the visibility of p and np is different
		if(abs(s - d.z) == 0.0) {
			//End the traversal
			shadowEdgeEnd = 1.0;
			break;
		//When the visibility of p and np is equal	
		} else {
			//Check if np and p have the same discontinuity dir.
			hasDiscontinuity = checkDiscontinuity(np, p, z, o);
			//Else, end the traversal
			if(!hasDiscontinuity) break;		
		}	
		//Increase the distance
		distance++;	
	}	
	
	return mix(-distance, distance, shadowEdgeEnd);	
	
}

float mix(float x, float y, float a) {
	return x * (1 - a) + y * a;
}
\end{lstlisting}

\subsubsection{Shadow Edge Traversal}

For every fragment inside a shadow edge, we need to search the ends of the shadow edge in order to estimate the size of the aliased shadow edge, as well as the relative distance of the fragment to the end of the shadow edge (Figure \ref{fig:SMSR}-(d)). This search is done in all the four directions of the 2D space (\textit{i.e.}, left, right, top and bottom directions), such that we can estimate the 2D relative position of the fragment in the shadow edge. The shadow edge traversal algorithm for each direction of the 2D space is implemented in Listing \ref{lst:DiscontinuitySpaceTraversal} by the function \lstinline{traverseShadowSilhouette}, that is called by \lstinline{computeFragmentDistanceToSE} of Listing \ref{lst:RBSM} for each search direction \lstinline{dir}.

For each shadow map neighbour of a given fragment (Lines 14-15 of Listing \ref{lst:DiscontinuitySpaceTraversal}), RBSM computes the shadow test (\ref{eq:ShadowTest}) for the neighbour (Lines 16-19 of Listing \ref{lst:DiscontinuitySpaceTraversal}) and detects whether the shadow test result of the neighbour is different from the one estimated for the given fragment (Lines 20-21 of Listing \ref{lst:DiscontinuitySpaceTraversal}). In this case, since the revectorization-based silhouette recovery algorithm operates only over lit fragments, a shadowed fragment has been detected. Therefore, we have detected the end of the shadow edge and we end the traversal in the particular direction (Lines 22-24 of Listing \ref{lst:DiscontinuitySpaceTraversal}). On the other hand, if the shadow test is the same between neighbour shadow map texels (Lines 25-26 of Listing \ref{lst:DiscontinuitySpaceTraversal}), we need to check if the neighbours share at least one discontinuity direction in common. If that is not the case, the neighbour shadow map texel accessed during traversal does not belong to the same shadow edge and the shadow traversal must be ended (Lines 27-30 of Listing \ref{lst:DiscontinuitySpaceTraversal}). To better understand this algorithm, let us visualize the scenario shown in Figure \ref{fig:SMSR}-(d). Only for the lit fragments inside the shadow edge, we perform the shadow edge traversal (resulting in the blue arrows depicted in Figure \ref{fig:SMSR}-(d)). To the right of these fragments, there are shadowed fragments that mark the end of the shadow edge. To the left of these fragments, there are other lit fragments. However, since these neighbour lit fragments does not have discontinuity directions, they do not belong to the same shadow edge.

To limit the extent of the shadow edge traversal, we define a variable \lstinline{MAXDIST}
which defines the maximum size of the shadow edge (Line 13 of Listing \ref{lst:DiscontinuitySpaceTraversal}). This variable is only used to improve the temporal consistency of the algorithm. Using \lstinline{MAXDIST = 16} was sufficient for our tests.

As a result of the shadow edge traversal, the algorithm returns the distance of the fragment to the end of the shadow edge. We orientate the distance value to be positive towards the end of the shadow edge (where the origin of the aliasing is located, as can be seen in Figure \ref{fig:SMSR}-(e)), and negative otherwise.

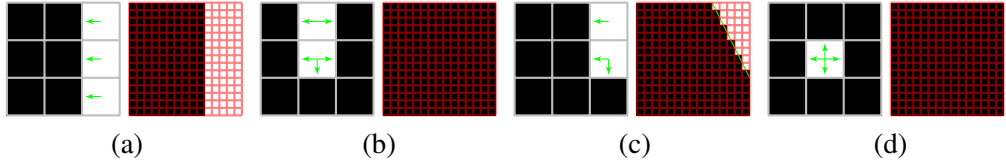
\begin{figure}[t!]
 	\centering
	\begin{tikzpicture}
	\tikzset {arrow/.style={->, >=latex', shorten >=1pt, ultra thick},}
	[>=stealth,line width=3, every node/.style={transform shape}]
		
		\begin{scope}[scale=0.5]
		\fill[black] (0, 0) rectangle (2, 3);
		\draw[arrow, very thin, color=green] (2.5, 0.5) -- (2.0, 0.5);
		\draw[arrow, very thin, color=green] (2.5, 1.5) -- (2.0, 1.5);
		\draw[arrow, very thin, color=green] (2.5, 2.5) -- (2.0, 2.5);
		\draw[thick, gray!50] (0, 0) grid (3, 3);
		
		\begin{scope}[shift={(3.25, 0)}]		
		\fill[black] (0, 0) rectangle (2, 3);
		\draw[thick, red, opacity=0.5, step=0.2cm] (0, 0) grid (3, 3);	
		\end{scope}		
		\end{scope}
		\node[align=center] at (1.6, -0.4) {(a)};
	
		\begin{scope}[scale=0.5, shift={(6.75, 0)}]
		\fill[black] (0, 0) rectangle (1, 3);
		\fill[black] (1, 0) rectangle (2, 1);
		\fill[black] (2, 0) rectangle (3, 3);
		\draw[arrow, very thin, color=green] (1.5, 2.5) -- (1.0, 2.5);
		\draw[arrow, very thin, color=green] (1.5, 2.5) -- (2.0, 2.5);
		\draw[arrow, very thin, color=green] (1.5, 1.5) -- (1.0, 1.5);
		\draw[arrow, very thin, color=green] (1.5, 1.5) -- (2.0, 1.5);
		\draw[arrow, very thin, color=green] (1.5, 1.5) -- (1.5, 1.0);
		\draw[thick, gray!50] (0, 0) grid (3, 3);
		
		\begin{scope}[shift={(3.25, 0)}]		
		\fill[black] (0, 0) rectangle (3, 3);
		\draw[thick, red, opacity=0.5, step=0.2cm] (0, 0) grid (3, 3);	
		\end{scope}		
		\end{scope}
		\node[align=center] at (4.95, -0.4) {(b)};
	
		\begin{scope}[scale=0.5, shift={(13.5, 0)}]
		\fill[black] (0, 0) rectangle (2, 3);
		\fill[black] (2, 0) rectangle (3, 1);
		\draw[arrow, very thin, color=green] (2.5, 1.5) -- (2.5, 1.0);
		\draw[arrow, very thin, color=green] (2.5, 1.5) -- (2.0, 1.5);
		\draw[arrow, very thin, color=green] (2.5, 2.5) -- (2.0, 2.5);
		\draw[thick, gray!50] (0, 0) grid (3, 3);
		
		\begin{scope}[shift={(3.25, 0)}]		
		\fill[black] (0, 0) rectangle (2, 3);
		\fill[black] (2, 0) rectangle (3, 1);
		\fill[black] (2, 1) rectangle (2.2, 2.8);	
		\fill[black] (2.2, 1) rectangle (2.4, 2.4);	
		\fill[black] (2.4, 1) rectangle (2.6, 2.0);	
		\fill[black] (2.6, 1) rectangle (2.8, 1.6);	
		\fill[black] (2.8, 1) rectangle (3.0, 1.2);	
		\draw[green] (2, 3) -- (3, 1);
		\draw[thick, red, opacity=0.5, step=0.2cm] (0, 0) grid (3, 3);	
		\end{scope}		
		\end{scope}
		\node[align=center] at (8.35, -0.4) {(c)};
		
		\begin{scope}[scale=0.5, shift={(20.25, 0)}]
		\fill[black] (0, 0) rectangle (1, 3);
		\fill[black] (1, 0) rectangle (2, 1);
		\fill[black] (1, 2) rectangle (2, 3);
		\fill[black] (2, 0) rectangle (3, 3);
		\draw[arrow, very thin, color=green] (1.5, 1.5) -- (1.0, 1.5);
		\draw[arrow, very thin, color=green] (1.5, 1.5) -- (2.0, 1.5);
		\draw[arrow, very thin, color=green] (1.5, 1.5) -- (1.5, 1.0);
		\draw[arrow, very thin, color=green] (1.5, 1.5) -- (1.5, 2.0);
		\draw[thick, gray!50] (0, 0) grid (3, 3);
		
		\begin{scope}[shift={(3.25, 0)}]		
		\fill[black] (0, 0) rectangle (3, 3);
		\draw[thick, red, opacity=0.5, step=0.2cm] (0, 0) grid (3, 3);	
		\end{scope}		
		\end{scope}
		\node[align=center] at (11.75, -0.4) {(d)};
	
	\end{tikzpicture}
	\caption{RBSM deals with four shadow edge shapes: (a) I-shape, (b) U-shape, (c) L-shape, and (d) O-shape. On the left of each sub-figure, we show the shadow edge produced by shadow mapping along with its discontinuity directions (green arrows). On the right, we show the expected silhouette recovery effect produced by RBSM.}
	\label{fig:SMSRShapes}
\end{figure}

\subsubsection{Shadow Edge Normalization}

After the computation of the distance of the fragment to the shadow edge, we need to orientate and normalize such value to the unit interval, as depicted in Figure \ref{fig:SMSR}-(e). The origin of this local coordinate system is located in the corner of the aliasing.

In this step, the algorithm not only normalizes the relative distance of the fragment to the shadow edge, but also identifies the type of the shadow edge. For instance, if during the shadow edge traversal, the algorithm finds two shadow edge ends for a particular axis, the fragment is located inside a U- or O-shaped edge (Figures \ref{fig:SMSRShapes}-(b, d)), because these are the only possible shadow edge shapes in which a lit fragment is located in between two shadowed neighbour fragments. On the other hand, if the algorithm does not find any shadow edge end for a particular axis, the fragment is located inside an I-shaped edge (Figure \ref{fig:SMSRShapes}-(a)). As we show in Section \ref{sec:VisibilityFunctions}, the normalized distance of the fragment to the shadow edge as well as the type of the shadow edge in which the fragment is located are essential information for the RBSM visibility functions.

\begin{figure}[ht]
\centering
  \includegraphics[width=0.975\linewidth]{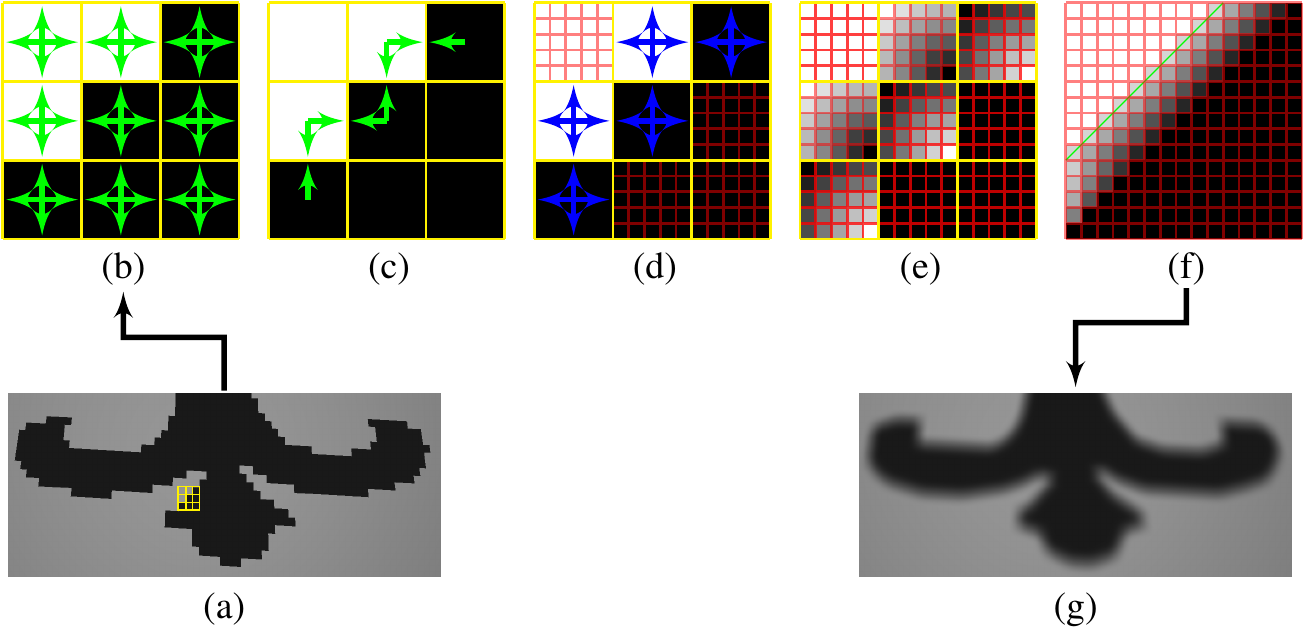}
  \caption{An overview of the RBSM pipeline for shadow filtering. Given an aliased shadow edge (a), RBSM evaluates the spatial coherency between shadow tests for shadowed and lit fragments (b), detects the directions (green arrows) of where the shadow silhouette is located (c), estimates the size of the aliased edge of the inner- and outer-side of the shadow edge (d) and computes the normalized distance, in the camera space (red grid), of each fragment, located in the aliased edge, to the origin of the local aliasing (e). Finally, the distance of each fragment to the revectorization line (green line in (f)) is calculated and used to determine the shadow intensity of each fragment (f, g).} 
  \label{fig:RSMSS}
\end{figure}

\subsection{Revectorization-Based Shadow Filtering}

\subsubsection{Overview}

RBSM can be used not only for shadow silhouette recovery, but also for shadow
filtering, producing fake penumbras of fixed size. An overview of the pipeline to produce revectorization-based shadow filtering is shown in Figure \ref{fig:RSMSS}.

As shown in Figure \ref{fig:RSMSS}-(b), one of the main differences between the silhouette recovery and the filtering variant of the RBSM is that the latter works over both inner- and outer-sides of the shadow edge to simulate the penumbra effect. So, the algorithm performs the neighbourhood evaluation (Figure \ref{fig:RSMSS}-(b)) for all the fragments visible in the scene, regardless of their initial illumination condition given by the shadow test. On the basis of the neighbourhood evaluation previously computed, the algorithm estimates the discontinuity directions for both sides of the shadow edge (Figure \ref{fig:RSMSS}-(c)). For each fragment located in the shadow edge, the algorithm performs a traversal over the shadow edge to estimate the size of the shadow edge and compute the relative distance of each fragment to the end of the shadow edge (Figure \ref{fig:RSMSS}-(d)), a value that is further normalized to the unit interval (Figure \ref{fig:RSMSS}-(e)) and used to define the final shadow intensity of each fragment (Figures \ref{fig:RSMSS}-(f, g)). It is noteworthy that, regardless of whether a fragment is located in the lit or shadowed part of shadow edge, the origin
of the local coordinate system of the aliased shadow edge remains the same (located at the corner of the aliasing, as can be seen in Figure \ref{fig:RSMSS}-(e)), such that the fragments are oriented towards this origin. 

In the next subsections, we present in more details how the implementation of
the shadow filtering variant of RBSM differs with respect to the revectorization-based shadow silhouette recovery variant.

\subsubsection{Shadow Edge Localization}

For the revectorization-based shadow filtering, we need to detect the directions where the shadow edge is located in both inner- and outer-sides of the shadow edge, because the shadow filtering covers both parts of the shadow edge, as shown in Figure \ref{fig:RSMSS}-(f).

For every fragment in the camera view, we compute the shadow test (\ref{eq:ShadowTest}), the neighbourhood evaluation (Figure \ref{fig:RSMSS}-(b) and (\ref{eq:NeighbourhoodEvaluation})) and the discontinuity directions (Figure \ref{fig:RSMSS}-(c) and (\ref{eq:compressedDiscontinuity})), as shown in Listings \ref{lst:RBSM} and \ref{lst:DiscontinuityComputation}. As a result of this step, we have the directions of where the shadow edge is located for all the fragments situated in the aliased shadow edge.

\subsubsection{Shadow Edge Traversal}

Similarly to the shadow silhouette recovery of RBSM, as shown in Listing \ref{lst:DiscontinuitySpaceTraversal}, during the traversal of the shadow edge, regardless of whether the fragment is located in the lit or shadowed side of the shadow edge, we still need to perform the shadow test for each neighbour shadow map texel being accessed (Lines 14-19 of Listing \ref{lst:DiscontinuitySpaceTraversal}). Then, when the neighbour shadow map texel has a different visibility condition than the initial shadow map texel, we have found the end of the shadow edge (Lines 20-24 of Listing \ref{lst:DiscontinuitySpaceTraversal}). Otherwise, if the neighbour shadow map texel has the same visibility condition of the initial shadow map texel of the traversal, we need to check whether these two point to the same discontinuity directions to determine whether the neighbour shadow map texel has stepped out of the aliased shadow edge (Lines 25-30 of Listing \ref{lst:DiscontinuitySpaceTraversal}).

\subsubsection{Shadow Edge Normalization}

As for the normalization of the relative distance computed in the previous step, we proceed similarly as the shadow silhouette recovery variant: we orientate and normalize the distance to the unit interval with respect to the corner of the aliased shadow edge. Also, we detect the type of the shadow edge in which the fragment is located at one of the shape patterns shown in Figure \ref{fig:SMSRShapes}. For the fragments located in the shadow, the shadow shape patterns are the same as the ones shown in Figure \ref{fig:SMSRShapes}. The difference is that, if the algorithm finds two shadow edge ends for a particular axis during the shadow edge traversal, the shadowed fragment is located inside an I-shaped edge (Figure \ref{fig:SMSRShapes}-(a)). On the other hand, if the algorithm does not find a shadow edge end for a specific axis, the shadowed fragment may be inside a U- or O-shaped edge (Figure \ref{fig:SMSRShapes}-(b, d)).

\section{Optimized RBSM Visibility Functions}
\label{sec:VisibilityFunctions}

As shown in Listing \ref{lst:RBSM}, the last step of RBSM is the definition of the visibility functions that will revectorize the shadow edge (Figures \ref{fig:SMSR}-(f, g) and \ref{fig:RSMSS}-(f, g)). Originally, RBSM handles 43 revectorization scenarios. 12 scenarios are handled for silhouette recovery (1 for I-shaped edges, 1 for O-shaped edges, 2 for U-shaped edges and 8 for L-shaped edges) and the other 31 scenarios are handled to produce the filtering effect (1 for I-shaped edges, 2 for O-shaped edges, 12 for U-shaped edges and 16 for L-shaped edges). Each one of these scenarios represents a different set of discontinuity directions located in the aliased shadow edge, and each revectorization solution for these scenarios is implemented as a specific branch in the shader code. So, the original RBSM solution is not only non-optimized, but also it makes difficult the understanding and implementation of the RBSM. Here, we propose a compact, symmetry-aware representation which takes advantage of the shadow edge shape to define the visibility functions. In this sense, we could reduce the 43 scenarios originally handled by RBSM to only 9 scenarios (4 for silhouette recovery and 5 for filtering). Each one of them related to a specific shadow edge shape. Hence, we first introduce an optimized version of the visibility function for silhouette recovery, then we introduce the optimized visibility function for filtering.

\begin{lstlisting}[caption={GLSL code for implementation of the silhouette recovery effect in RBSM.}, label={lst:SMSR}, float]
float computeRBSMVisibility(vec3 d, vec4 nrd) {

	//If short U- or O-shape, return 0.0 (shadow)
	if(d.x == 0.75 || d.y == 0.75) return 0.0;
	//If long U- or O-shape, return 0.0 (shadow)
	if(nrd.z == 1.0 || nrd.w == 1.0) return 0.0;
	//If I-shape, return 1.0 (lit)
	if(nrd.z == -1.0 || nrd.w == -1.0) return 1.0;
	//If L-shape, revectorize shadow edge
	return step(1.0 - nrd.x, nrd.y);
	
}

float step(float edge, float x) {
	return (x < edge) ? 0.0 : 1.0;
}

\end{lstlisting}

\subsection{Silhouette Recovery Visibility Function}

As shown in Listing \ref{lst:SMSR}, to perform the shadow revectorization, each fragment has to access only two variables: the discontinuity directions (variable \lstinline{d}) and the normalized distances to the origin of the aliased shadow edge (variable \lstinline{nrd}). As a \lstinline{vec4}, \lstinline{nrd} stores not only the normalized distances of the fragment in its first two components (\lstinline{nrd.x}, \lstinline{nrd.y}), but also an indication if a shadow edge end was found during the traversal of both directions of a particular axis in its latter two components (\lstinline{nrd.z}, \lstinline{nrd.w}). This indication is labelled as follows

\newsavebox\boxNZW
\savebox\boxNZW{\lstinline{nrd.zw}}

\begin{equation}
\label{eq:NRD}
\usebox{\boxNZW} = 
\begin{cases}
	1 & \text{if a shadow edge end was found for both directions,}
	\\
	0 & \text{if a shadow edge end was found in only one direction},
	\\
	-1 & \text{if none shadow edge end was found}.
\end{cases}
\end{equation}

In (\ref{eq:NRD}), the component \lstinline{nrd.z} refers to the horizontal axis, meanwhile the component \lstinline{nrd.w} refers to the vertical axis.

\begin{figure}[ht!]
\centering
  \includegraphics[width=0.975\linewidth]{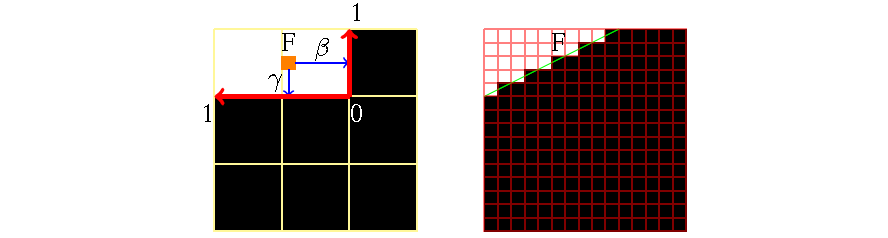}
  \caption{By using the normalized distances ($\gamma$ and $\beta$) of the fragment (F) to the shadow edge, RBSM is able to define a revectorized line (green line) which defines the visibility condition of the fragment (b).} 
  \label{fig:ONDSClipping}
\end{figure}

For silhouette recovery, we deal with the four shadow edge shapes shown in Figure \ref{fig:SMSRShapes}. U- and O-shaped edges can be easily identified when the discontinuity directions are pointing out two opposite directions of the same axis (Figures \ref{fig:SMSRShapes}-(b, d) and Lines 3-4 of Listing \ref{lst:SMSR}), or when a shadow edge end was found for both sides of the same axis (Lines 5-6 of Listing \ref{lst:SMSR}). In both cases, we follow the original RBSM visibility function and put the fragment in shadow, producing a visual effect similar to the one depicted in Figure \ref{fig:SMSRShapes}-(b, d). A shadow edge with the I-shape (Figure \ref{fig:SMSRShapes}-(a)) can be found when none shadow edge end was found for both sides of the same axis (Lines 7-8 of Listing \ref{lst:SMSR}). In this case, we maintain the fragment as lit, keeping the visual effect shown in Figure \ref{fig:SMSRShapes}-(a).

To anti-alias the L-shaped shadow edge, we fit a revectorization line where all fragments belonging to the interior of this line are put in shadow (green line in Figure \ref{fig:ONDSClipping}). This revectorization line results from a linear comparison between the normalized distances of the fragment to the shadow edge for both horizontal and vertical axes. Given the normalized coordinate system shown in Figure \ref{fig:ONDSClipping}, a fragment is shadowed by the shadow revectorization technique if the vertical distance of the fragment to the shadow edge is lower than 1 minus the horizontal distance of the fragment to the shadow edge. Otherwise, the fragment remains lit. This visibility function can also be seen in Lines 9, 10 of Listing \ref{lst:SMSR}, where the \lstinline{step} function is used for the linear comparison.

\subsection{Filtering Visibility Function}

For filtering, we deal with the same four basic shadow edge shapes shown in Figure \ref{fig:SMSRShapes}. However, we handle discontinuities located in both the interior and exterior side of the shadow edge. The original RBSM handles a set of 31 visibility functions to perform revectorization-based shadow filtering. Differently from the silhouette recovery approach, where a linear comparison is used to determine the binary visibility condition of a fragment, for filtering, RBSM uses addition and subtraction operations to determine the floating-point visibility condition of the fragment.

\begin{lstlisting}[caption={GLSL code for implementation of the filtering effect in RBSM.}, label={lst:RSMSS}, float]
float computeRBSMVisibility(vec3 d, vec4 nrd) {

	float sign = -2.0 * d.z + 1.0;
	//If short O-shape, return the opposite of the shadow test
	if(d.x == 0.75 && d.y == 0.75) return d.z;
	//If long O-shape, return the opposite of the shadow test
	if(nrd.z == 1.0 && nrd.w == 1.0) return d.z;
	//If U-shape, return the normalized distance
	if(nrd.z == 1.0) return d.z + sign * nrd.y;
	if(nrd.w == 1.0) return d.z + sign * nrd.x;	
	//If I-shape, return the shadow test	
	if(nrd.z == -1.0 || nrd.w == 1.0) return 1.0 - d.z;
	//If L-shape, filter shadow edge
	return clamp(d.z + sign * nrd.y + sign * nrd.x, 0.0, 1.0);
	
}

\end{lstlisting}

For O-shaped edges (Figure \ref{fig:SMSRShapes}-(d)), we simply return the opposite of the shadow test to close the shadow edge (Lines 4-7 of Listing \ref{lst:RSMSS}). For U-shaped edges (Figure \ref{fig:SMSRShapes}-(b)), we return the normalized distance of the fragment to the shadow edge in order to simulate the filtering effect (Lines 8-10 of Listing \ref{lst:RSMSS}). For I-shaped edges (Figure \ref{fig:SMSRShapes}-(a)), we simply return the shadow test (Lines 11-12 of Listing \ref{lst:RSMSS}). Finally, for L-shaped edges, depending on the side of shadow edge in which the fragment is located, an addition or subtraction operation is computed with the normalized distances of the fragment to the shadow edge (Lines 13-14 of Listing \ref{lst:RSMSS}).

\section{Results and Discussion}
\label{sec:Results}

\begin{figure}[t!]
\centering
	\begin{tikzpicture}[>=stealth,line width=3, every node/.style={transform shape}, spy using outlines={rectangle,red,magnification=3.8, width=6.3cm, height=1.75cm}]
	
	\node [scale=0.19] at (0, 0) {\pgfuseimage{Fence-SM}};
	\node [scale=0.19] at (6.4, 0) {\pgfuseimage{Fence-RBSM}};
	\node [scale=0.19] at (3.2, -6.0) {\pgfuseimage{Fence-SV}};
	
	\spy [opacity=1.0] on (-0.05, -1.025) in node [left, opacity=0.0] at (3.15, -2.7);
	\spy [opacity=1.0] on (6.35, -1.025) in node [left, opacity=0.0] at (3.15 + 6.4, -2.7);
	\spy [opacity=1.0] on (3.15, -7.025) in node [left, opacity=0.0] at (3.15 + 3.2, -8.7);
	
	\node at (0, -4.0) {(a) Shadow Mapping};
	\node at (6.4, -4.0) {(b) Optimized RBSM};
	\node at (3.2, -10.0) {(c) Shadow Volume};
		
	\end{tikzpicture}
	\caption{A visual comparison between different shadow silhouette recovery techniques. This Fence scene was rendered using a $2048^{2}$ shadow map resolution.} 
  \label{fig:SMSRVisualQuality}
\end{figure}
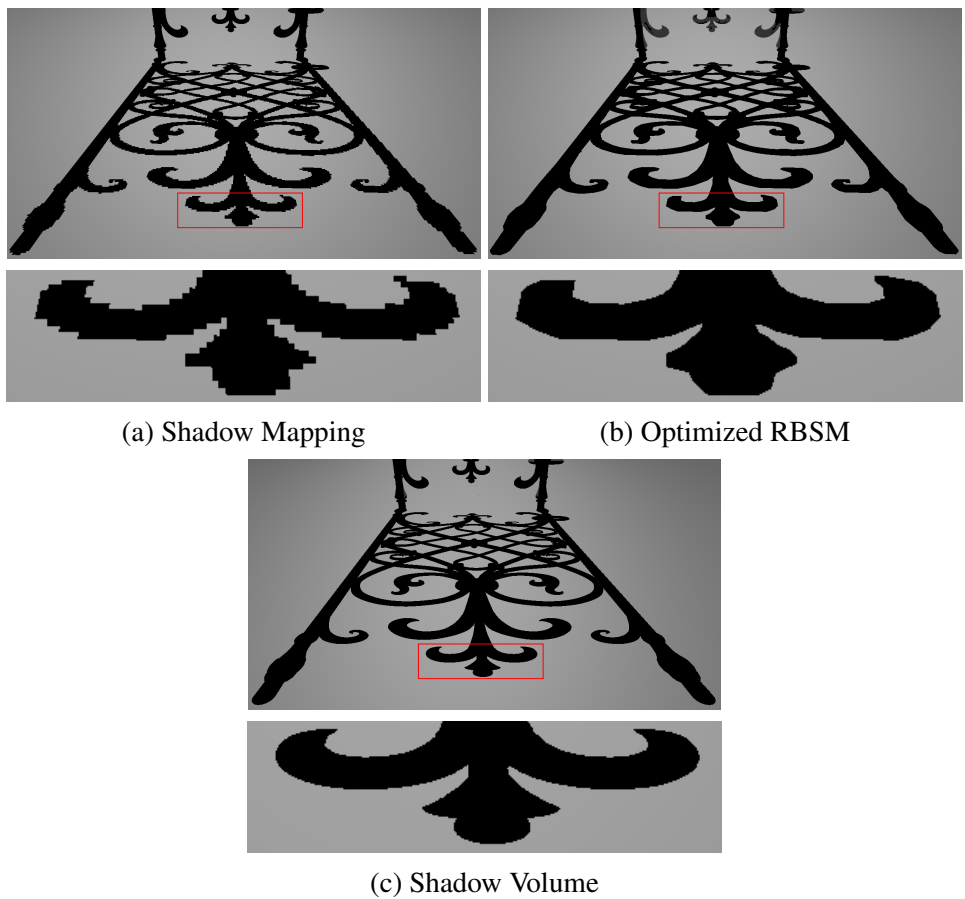

\begin{figure}[t!]
\centering
	\begin{tikzpicture}[>=stealth,line width=3, every node/.style={transform shape}, spy using outlines={rectangle,magnification=4.0, width=3.0cm, height=1.75cm}]
	
	\node [scale=0.25] at (0, 0) {\pgfuseimage{QuadBot-PCF}};
	\node [scale=0.25] at (6.4, 0) {\pgfuseimage{QuadBot-VSM}};
	\node [scale=0.25] at (0.0, -6.25) {\pgfuseimage{QuadBot-MSM}};
	\node [scale=0.25] at (6.4, -6.25) {\pgfuseimage{QuadBot-RBSM}};
	
	\spy [red, opacity=1.0] on (-2.4, -1.3) in node [left, red, opacity=1.0] at (-0.1, -2.85);
	\spy [green, opacity=1.0] on (-2.0, 0.0) in node [left, green, opacity=1.0] at (3.1, -2.85);
	\spy [red, opacity=1.0] on (-2.4 + 6.4, -1.3) in node [left, red, opacity=1.0] at (-0.1 + 6.4, -2.85);
	\spy [green, opacity=1.0] on (-2.0 + 6.4, 0.0) in node [left, green, opacity=1.0] at (3.1 + 6.4, -2.85);
	\spy [red, opacity=1.0] on (-2.4, -1.3 - 6.25) in node [left, red, opacity=1.0] at (-0.1, -2.85 - 6.25);
	\spy [green, opacity=1.0] on (-2.0, 0.0 - 6.25) in node [left, green, opacity=1.0] at (3.1, -2.85 - 6.25);
	\spy [red, opacity=1.0] on (-2.4 + 6.4, -1.3 - 6.25) in node [left, red, opacity=1.0] at (-0.1 + 6.4, -2.85 - 6.25);
	\spy [green, opacity=1.0] on (-2.0 + 6.4, 0.0 - 6.25) in node [left, green, opacity=1.0] at (3.1 + 6.4, -2.85 - 6.25);
	
	\node at (0, -4.15) {(a) PCF};
	\node at (6.4, -4.15) {(b) VSM};
	\node at (0.0, -4.15 -6.25) {(c) MSM};
	\node at (6.4, -4.15 -6.25) {(d) Optimized RBSM};
		
	\end{tikzpicture}
	\caption{A visual comparison between different shadow filtering techniques. This QuadBot scene was rendered using a $1024^{2}$ shadow map resolution.} 
  \label{fig:RSMSSVisualQuality}
\end{figure}
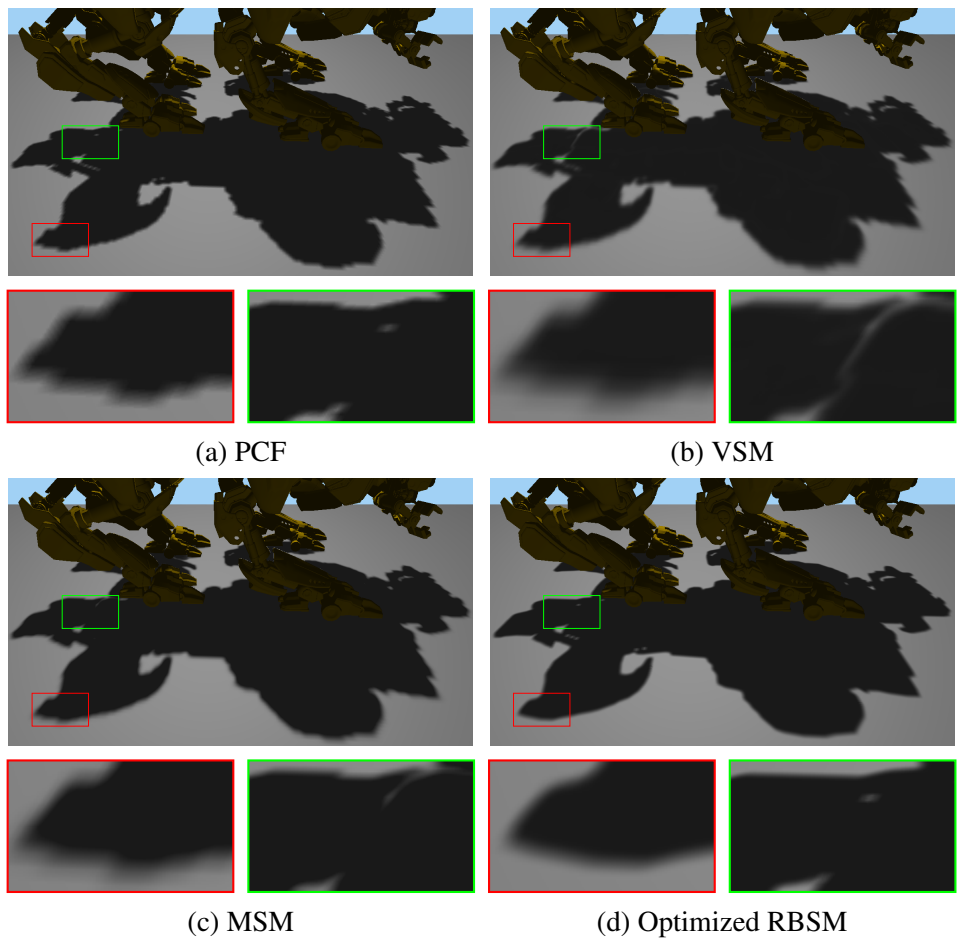

In this section, we evaluate the optimized implementation of RBSM mainly in terms of performance and visual quality. In our experimental setup, processing time and visual quality were evaluated in an Intel Core\textsuperscript{TM} i7-3770K CPU (3.50 GHz), 8GB RAM, and an NVIDIA GeForce GTX Titan X graphics card. For the silhouette recovery variant of RBSM, we compare the optimized approach with the shadow volume algorithm proposed in \cite{Heidmann1991}. For the filtering variant, we compare the proposed approach with the traditional Percentage-Closer Filtering (PCF) \cite{Reeves1987} and Variance Shadow Mapping (VSM) \cite{Donnelly2006} algorithms and the recent Moment Shadow Mapping (MSM) \cite{Peters2015}.

\subsection{Visual Quality}

In Figure \ref{fig:SMSRVisualQuality}, we compare the hard shadows generated with shadow mapping, RBSM and the shadow volume technique, the latter a reference technique for accurate hard shadow generation. Shadow mapping produces aliasing artifacts along the shadow edge (Figure \ref{fig:SMSRVisualQuality}-(a)) even with the use of a high-resolution shadow map. On the other hand, RBSM minimizes the aliasing artifacts along the shadow edge, recovering an approximate accurate shadow edge (Figure \ref{fig:SMSRVisualQuality}-(b)). Finally, the shadow volume technique produces the most accurate shadows, capturing fine details of the shadow edge that are missed by both shadow mapping and RBSM techniques (Figure \ref{fig:SMSRVisualQuality}-(c)).

In Figure \ref{fig:RSMSSVisualQuality}, we compare different shadow filtering techniques with respect to the generation of aliasing (red closeups in Figure \ref{fig:RSMSSVisualQuality}) and light leaking artifacts (green closeups in Figure \ref{fig:RSMSSVisualQuality}). As shown in the red closeup of Figure \ref{fig:RSMSSVisualQuality}-(a), the PCF technique is able to generate fake penumbras, but is prone to aliasing artifacts along the shadow edge. The VSM technique minimizes the aliasing artifacts (red closeup of Figure \ref{fig:RSMSSVisualQuality}-(b)), but is prone to light leaking artifacts (green closeup of Figure \ref{fig:RSMSSVisualQuality}-(b)). As a direct improvement of VSM, MSM is able to minimize the light leaking artifacts (green closeup of Figure \ref{fig:RSMSSVisualQuality}-(c)). RBSM does not suffer from light leaking artifacts as much as VSM and MSM, and can suppress the aliasing artifacts (Figure \ref{fig:RSMSSVisualQuality}-(d)), improving the shadow visual quality.

\begin{table}[t]
\centering
\begin{tabular}{c|c|c|c|c|}
\cline{2-5}
\multicolumn{1}{c|}{}                 & \multicolumn{4}{c|}{\textbf{Shadow Map Resolution}} \\ \hline
\multicolumn{1}{|c|}{\textbf{Method}} & $512^{2}$   & \multicolumn{1}{c|}{$1024^{2}$}   & $2048^{2}$  & $4096^{2}$  \\ \hline
\multicolumn{1}{|c|}{Shadow Mapping}  & 3.03 ms & 3.05 ms & 3.07 ms & 3.30 ms \\ \hline
\multicolumn{1}{|c|}{RBSM (Optimized)} & 3.22 ms & 3.32 ms & 3.49 ms & 4.34 ms \\ \hline
\multicolumn{1}{|c|}{RBSM (Non-Optimized)} & 3.23 ms & 3.33 ms & 3.51 ms & 4.36 ms \\ \hline
\multicolumn{1}{|c|}{Shadow Volumes}  & 140.20 ms & 140.20 ms & 140.20 ms & 140.20 ms \\ \hline
\end{tabular}
\caption{Performance results measured for different silhouette recovery techniques for a $1280 \times 720$ output resolution. Measurement include varying shadow map resolution for the Fence scene shown in Figure \ref{fig:SMSRVisualQuality}.}
\label{tab:SMSRShadowMapResolution}
\end{table}

\begin{table}[t]
\centering
\begin{tabular}{c|c|c|c|}
\cline{2-3}
\multicolumn{1}{c|}{}                 & \multicolumn{2}{c|}{\textbf{Viewport Resolution}} \\ \hline
\multicolumn{1}{|c|}{\textbf{Method}} & $1280 \times 720$   & \multicolumn{1}{c|}{$1920 \times 1080$}   \\ \hline
\multicolumn{1}{|c|}{Shadow Mapping}  & 3.07 ms & 3.67 ms \\ \hline
\multicolumn{1}{|c|}{RBSM (Optimized)} & 3.49 ms & 4.29 ms \\ \hline
\multicolumn{1}{|c|}{RBSM (Non-Optimized)} & 3.51 ms & 4.31 ms\\ \hline
\multicolumn{1}{|c|}{Shadow Volumes}  & 140.20 ms & 297.62 ms\\ \hline
\end{tabular}
\caption{Performance results measured for different silhouette recovery techniques for a $2048^{2}$ shadow map resolution. Measurement include varying output resolution for the Fence scene shown in Figure \ref{fig:SMSRVisualQuality}.}
\label{tab:SMSROutputResolution}
\end{table}

\begin{table}[t]
\centering
\begin{tabular}{c|c|c|c|c|}
\cline{2-5}
\multicolumn{1}{c|}{}                 & \multicolumn{4}{c|}{\textbf{Shadow Map Resolution}} \\ \hline
\multicolumn{1}{|c|}{\textbf{Method}} & $512^{2}$   & \multicolumn{1}{c|}{$1024^{2}$}   & $2048^{2}$  & $4096^{2}$  \\ \hline
\multicolumn{1}{|c|}{Shadow Mapping}  & 5.49 ms & 5.53 ms & 5.58 ms & 5.98 ms \\ \hline
\multicolumn{1}{|c|}{PCF}  & 6.25 ms & 6.75 ms & 6.82 ms & 7.40 ms \\ \hline
\multicolumn{1}{|c|}{VSM}  & 6.45 ms & 6.81 ms & 7.14 ms & 8.00 ms \\ \hline
\multicolumn{1}{|c|}{MSM}  & 6.53 ms & 6.84 ms & 7.09 ms & 8.06 ms \\ \hline
\multicolumn{1}{|c|}{RBSM (Optimized)} & 7.04 ms & 7.70 ms & 8.14 ms & 9.40 ms \\ \hline
\multicolumn{1}{|c|}{RBSM (Non-Optimized)} & 7.07 ms & 7.73 ms & 8.20 ms & 9.46 ms \\ \hline
\end{tabular}
\caption{Performance results measured for different filtering techniques for a $1280 \times 720$ output resolution. Measurement include varying shadow map resolution for the QuadBot scene shown in Figure \ref{fig:RSMSSVisualQuality}.}
\label{tab:RSMSSShadowMapResolution}
\end{table}

\begin{table}[t]
\centering
\begin{tabular}{c|c|c|c|}
\cline{2-3}
\multicolumn{1}{c|}{}                 & \multicolumn{2}{c|}{\textbf{Viewport Resolution}} \\ \hline
\multicolumn{1}{|c|}{\textbf{Method}} & $1280 \times 720$   & \multicolumn{1}{c|}{$1920 \times 1080$} \\ \hline
\multicolumn{1}{|c|}{Shadow Mapping}  & 5.53 ms & 6.25 ms \\ \hline
\multicolumn{1}{|c|}{PCF}  & 6.75 ms & 7.40 ms \\ \hline
\multicolumn{1}{|c|}{VSM}  & 6.81 ms & 8.13 ms \\ \hline
\multicolumn{1}{|c|}{MSM}  & 6.84 ms & 8.00 ms \\ \hline
\multicolumn{1}{|c|}{RBSM (Optimized)} & 7.70 ms & 9.98 ms \\ \hline
\multicolumn{1}{|c|}{RBSM (Non-Optimized)} & 7.73 ms & 10.04 ms \\ \hline
\end{tabular}
\caption{Performance results measured for different filtering techniques for a $1024^{2}$ shadow map resolution. Measurement include varying output resolution for the QuadBot scene shown in Figure \ref{fig:RSMSSVisualQuality}.}
\label{tab:RSMSSOutputResolution}
\end{table}

\subsection{Rendering Time}

In Tables \ref{tab:SMSRShadowMapResolution} and \ref{tab:SMSROutputResolution}, we show a performance comparison between the silhouette recovery techniques for different shadow map resolutions and output window sizes. Shadow volume generates the most accurate shadows, but is about two orders of magnitude slower than both shadow mapping and RBSM techniques. Meanwhile, the silhouette recovery variant of RBSM is as fast as shadow mapping. Our simpler, optimized implementation of RBSM is nearly 1\% faster than previous RBSM implementation. 

In Tables \ref{tab:RSMSSShadowMapResolution} and \ref{tab:RSMSSOutputResolution}, we provide a rendering time comparison between different shadow filtering techniques. Even when using high shadow map resolutions ($2048^{2}$ and $4096^{2}$), all the techniques evaluated in this section require less than 10 milliseconds to compute filtered hard shadows. RBSM is able to generate the most visually pleasant fake penumbras, nevertheless, it is 10-20\% slower than related work for low- and high- viewport resolution (Table \ref{tab:RSMSSOutputResolution}), and different shadow map resolutions (Table \ref{tab:RSMSSShadowMapResolution}).


\begin{figure}[t!]
\centering
	\begin{tikzpicture}[>=stealth,line width=0.5, every node/.style={transform shape}]
	
	\node [label=below:(a)][scale=0.28] at (0, 0) {\pgfuseimage{Limitations-SMSR-SM-1024}};
	\node [label=below:(b)][scale=0.28] at (4.4, 0) {\pgfuseimage{Limitations-SMSR-SM-2048}};
	\node [label=below:(c)][scale=0.28] at (8.8, 0) {\pgfuseimage{Limitations-SMSR-SM-4096}};
	\node [label=below:(d)][scale=0.28] at (0.0, -4.8) {\pgfuseimage{Limitations-SMSR-SMSR-1024}};
	\node [label=below:(e)][scale=0.28] at (4.4, -4.8) {\pgfuseimage{Limitations-SMSR-SMSR-2048}};
	\node [label=below:(f)][scale=0.28] at (8.8, -4.8) {\pgfuseimage{Limitations-SMSR-SMSR-4096}};
	
	\draw [red] (0.1, -3.5) rectangle (0.4, -3.25);
	\draw [red] (4.5, -3.5) rectangle (4.8, -3.25);
	\draw [red] (8.9, -3.5) rectangle (9.2, -3.25);
			
	\draw [red] (-1.65, -3.9) rectangle (-1.3, -3.65);
	\draw [red] (4.4 - 1.65, -3.9) rectangle (4.4 - 1.3, -3.65);
	\draw [red] (8.8 - 1.65, -3.9) rectangle (8.8 - 1.3, -3.65);
	
	\draw [red] (-1.45, -4.55) rectangle (-0.95, -4.15);
	\draw [red] (4.4 - 1.45, -4.55) rectangle (4.4 - 0.95, -4.15);
	\draw [red] (8.8 - 1.45, -4.55) rectangle (8.8 - 0.95, -4.15);
	
	\end{tikzpicture}
	\caption{A comparison between shadow mapping (a, b, c) and RBSM (d, e, f) using $1024^{2}$ (a, d), $2048^{2}$ (b, e), $4096^{2}$ (c, f) shadow map resolution. Some artifacts may arise due to the revectorization on the jagged shadow edges produced by the low shadow map resolution (red rectangles).} 
  \label{fig:LimitationsSMSR}
\end{figure}
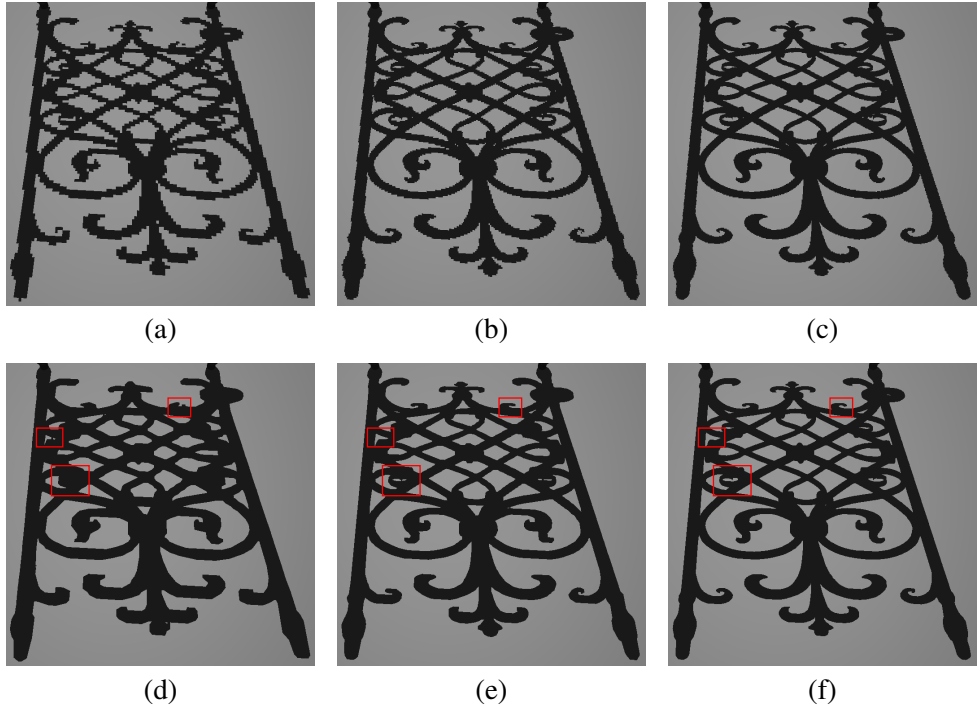

By designing the visibility functions according to the shadow edge shape, we could simplify the implementation of the RBSM technique. So, our approach is not only faster than the previous implementation of RBSM, but is also simpler to implement than the original RBSM. For filtering, we could optimize the original RBSM, however, the approach is still more costly than the silhouette recovery technique.

\subsection{Limitations}

The main limitation of the silhouette recovery effect produced by shadow revectorization relies on its accuracy, which is highly dependent on the shadow map resolution. So, for insufficient shadow map resolutions, the shadow revectorization minimizes aliasing, but can introduce shadow overestimation (Figure \ref{fig:LimitationsSMSR}-(a, d)). For high-resolution shadow maps, which are still prone to aliasing artifacts because of the finite resolution of the shadow map, revectorization effectively suppresses the aliasing artifacts and improves the visual quality of the shadowed scene (Figure \ref{fig:LimitationsSMSR}-(b, c, e, f)).


\begin{figure}[t!]
\centering
	\begin{tikzpicture}[>=stealth,line width=0.5, every node/.style={transform shape}]
	
	\node [label=below:(a)][scale=0.2] at (0, 0) {\pgfuseimage{Limitations-RSMSS-SM}};
	\node [label=below:(b)][scale=0.2] at (6.0, 0) {\pgfuseimage{Limitations-RSMSS-RSMSS}};
	
	\end{tikzpicture}
	\caption{The RBSM visibility function for filtering does not handle some shadow edge shapes produced by shadow mapping.} 
  \label{fig:LimitationsRSMSS}
\end{figure}
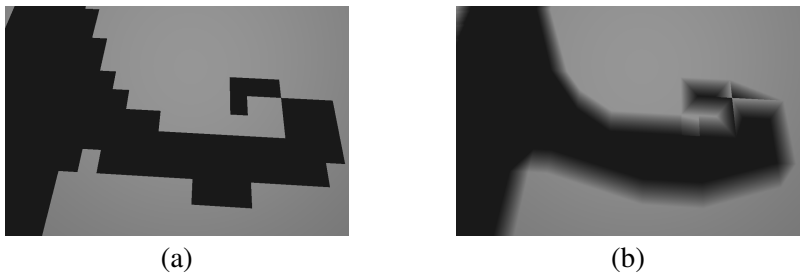

As for the filtering effect, the RBSM visibility function is able to handle the most common shadow edge shapes, however, it produces unpleasant results for some unhandled scenarios (Figure \ref{fig:LimitationsRSMSS}). As the handling of such scenarios is costly, because one would need to access several shadow map samples to identify the type of such a specific shadow edge shape, the proposal of an efficient visibility function to handle all the possible shadow edge shapes remains open. Moreover, according to the definition of the visibility function, skeleton artifacts appear along the filtered shadow edges. These can be suppressed by the application of PCF \cite{Reeves1987}.

\section{Final Remarks}
\label{sec:Conclusion}

In this paper, we have shown an algorithm that solves the shadow aliasing by revectorizing shadow edges. We have presented implementation details of the original RBSM algorithm, as well as our optimized visibility functions, which runs about 1\% faster than the original technique, while being easier to implement. 

For future work, one can further investigate if the RBSM visibility functions can be even more optimized. Also, an integration and evaluation of RBSM in the existing game engines would help on the popularization of the technique. Finally, a hybrid approach between RBSM and related work (such as \cite{Lecocq2014}) could be proposed to improve the accuracy of the revectorization. 

\subsection*{Acknowledgements}

The authors would like to thank Coordena\c{c}\~ao de Aperfei\c{c}oamento de Pessoal do N\'ivel Superior (CAPES) and NVIDIA for donating the hardware used to evaluate the techniques through the GPU Education Center program. Fence model is courtesy of Archive3D (user Nike). 

\small
\bibliographystyle{jcgt}
\bibliography{references}

\end{document}